\documentclass[12pt]{iopart}
\usepackage{iopams}  
\usepackage{amstext}
\usepackage{graphicx}

\renewcommand\vec[1]{\boldsymbol{#1}}

\begin{document}

\title{A simple model of the charge transfer in DNA-like substances}

\author{Niels R. Walet\dag \footnote[3]{Niels.Walet@manchester.ac.uk}
 and Wojtek J. Zakrzewski\ddag \footnote[4]{W.J.Zakrzewski@durham.ac.uk}}

\address{\dag\ School of Physics and Astronomy, University of Manchester, P.O. Box 88, Manchester M60 1QD, UK}
\address{\ddag\ Dept. of Mathematical Sciences, University of Durham, Durham DH1
3LE, UK}

\begin{abstract}
We present a very simple model for the study of charge transport in
a molecule patterned on B-DNA. In this model we use a discrete non-linear
Schr\"{o}dinger equation to describe electrons propagating along
the sugar-phosphate backbone of the DNA molecule. We find that in
this model, for a given nonlinearity, the transport is controlled
by $J$, a parameter which relates to the electronic coupling between
the different molecules on the backbone. For smaller values of $J$ we
have localised states while at higher values of $J$ the soliton field
is spread out and through its interaction with the lattice it has
stronger effects on the distortion of the lattice. 
\end{abstract}
\maketitle

\section{introduction}

As the molecule that carries all genetic information, DNA has been of
great interest to chemists and physicists since before the unlocking of
its structure by Watson and Crick \cite{WatsonCrick53}, and both the
structure and dynamics of the molecule have been investigated in great
detail. Another key feature of DNA is the ease of its assembly as
a linear macro-molecule of truly macroscopic scale; a potential
realisation of a quantum wire (if it were a conductor).  One can also
use the DNA molecule as a scaffold to grow quantum wires, e.g., by
deposition of conducting molecules in the minor
groove.\cite{RichterMPMS01,JoshiT01,AichSWSL02,Richter03}

With the experiments by Dekker and his group \cite{PorathBdVD00,StormvdD01,CunibertiCPD02}
on the conduction of single strands of DNA, and the controversy about
the interpretation of the experiments, which suggested that DNA might
be a conductor, a veritable floodgate for theoretical studies has
opened. Most of these consider the conduction (or charge transport)
along a one-dimensional {}``pi-stack'' (a channel built from overlapping
pi orbitals) in the centre of the DNA molecule.

Disregarding electronic transport, mathematical modelling of DNA has a
long tradition, see for example the book Ref. \cite{Yakush98}. In
recent years the Peyrard-Bishop model (a simple two-dimensional model
for DNA, which has been used successfully to describe the denaturing
transition of that molecule) has become quite popular, see the review
in \cite{Peyrard04} for a more complete set of references. Mingaleev
\emph{et al} \cite{MingaleevCGJR99,Christiansen01} have
produced models which allow them to perform mathematical studies of
DNA and its properties. This is an active field, and the number of
relevant publications is large. The most relevant literature can be traced
from a number of recent references,
e.g., Refs. \cite{Agarwal03,Cuevas03,Dong03,Hadzievski03,Hennig03,Iguchi03,Kalosakas03,Kevrekidis03,Kevrekidis03a,Maniadis03,Papacharalampo03} and
the review \cite{ECS04}.

Recently one of the present authors (together with Dr. Hartmann)
\cite{HartmannZ02} looked at the model of Mingaleev \emph{et al}
and proposed a further generalisation of their one chain model to two
chains. This two chain generalisation involved similar terms in the
Lagrangian (for each chain) to those of the model of Mingaleev \emph{et
al} plus an extra term involving the interaction between the chains.
 The obtained results were encouraging, although due
to the specific choice of terms, the numerical simulations were very
slow. 

In this note we reinvestigate these models and propose a new
generalisation which, at least, has the advantage of fast numerical
convergence. We use this model to investigate various aspects of DNA-like
substances. Our model can be compared to the model used by Bishop \emph{et al.}
\cite{Maniadis03}, but differs in many aspects. The two most crucial
ones are the inclusion of the helicity of the molecule 
and the fact that in our model electrons are (correctly or incorrectly, that
remains to be seen) allowed to propagate along the backbone rather
than along the {}``pi-stack'' of the overlapping molecular orbitals
of the base pairs. The latter assumption is already quite old, and
was first proposed by Eley \cite{EleyS62} in 1962. In this work we
concentrate our attention on the mathematical aspects of the problem
postponing detailed aspects of the applicability of the model
to real, physical and biological, systems to a further publication.

In the next section we say a few words about the DNA systems and in
the following section we describe our ideas of how to build a model
for DNA-inspired structures which can be used to describe such systems.

In the final section we describe in detail our first studies involving
this model.

\section{Some aspects of DNA}

\begin{figure}
\begin{center}\includegraphics[height=10cm,  keepaspectratio]{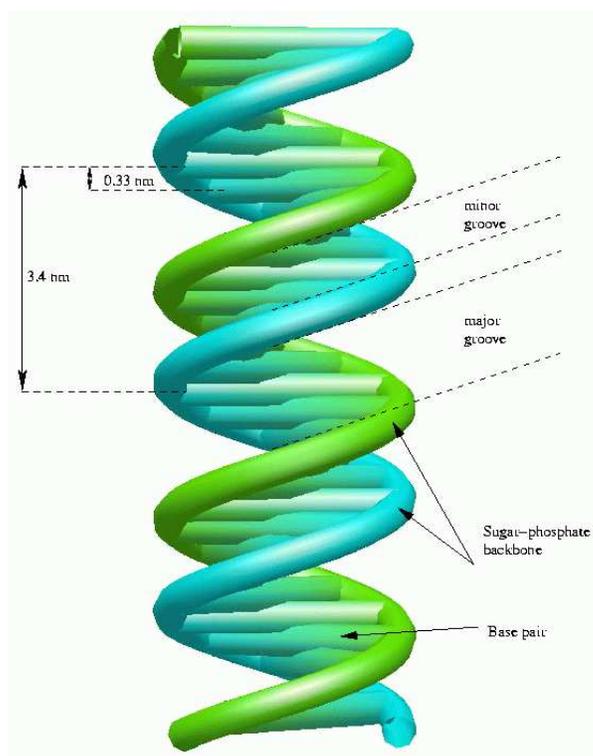}\end{center}

\caption{\label{fig:BDNA}A schematic model of B-DNA}
\end{figure}

The basic structure of DNA varies quite considerably depending on
its environment. B-DNA, the standard double helix, actually only occurs
in aqueous solution. An idealised structure of DNA is given in figure
\ref{fig:BDNA}.

We can easily see that each base pair advances the rotation of the
chain by about $36^{\circ}$. The difference between the major and
minor grooves is caused by the fact that the base-pairs are not exactly
at right angles to the backbone. There is also a {}``propeller twist'':
the plane of the base pairs is twisted like a propeller. See Ref.~\cite{CallandineD97}
for more details. As can be seen in the figure, the different base
pairs are quite close in the centre of the molecule. This is why it
is believed that DNA gains rigidity by overlapping pi-orbitals between
the different base pairs. This same mechanism has been invoked in
allowing charge transport through DNA.

\section{Modelling DNA}

\begin{figure}
\begin{center}\includegraphics[%
  height=3in,
  keepaspectratio]{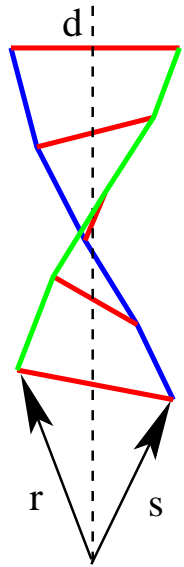}\end{center}

\caption{\label{fig:DNA1}A simple model of a section of straight DNA}
\end{figure}

Our DNA model will consist of two parts: a simple classical model
of a chiral molecule, that can twist and be locally compressed, but
where we disregard the bending modes. This would be a natural approach to a stretched DNA molecule, attached between two electrical contacts.
The situation is sketched schematically in figure~\ref{fig:DNA1}
\footnote{This may not be the best approach; stretching of the {}``rungs''
may well be ignored at low temperatures, but the bending of the
backbone seems to be a low-energy degree of freedom. For simplicity we
do not model the very complicated bending and kinking
\cite{Levitt83}.}.  In that figure the thin (red) lines denote denote
the hydrogen bonds, and are modelled to be at straight angles to the
{}``virtual backbone'' denoted by the dashed line. The thick (blue and
green) lines denote the two DNA chains.

At the same time we assume electrons hop from a base pair to a base pair,
occasionally moving from one side to the other. Our model for this
process will be based on the discrete non-linear Schr\"odinger equation,
which can be seen as the semi-classical limit of a tight-binding model.

We shall investigate the electronic model first, before concentrating
on the molecular dynamics.

\subsection{Electron dynamics}
We assume a simple discrete non-linear Schr\"odinger equation for the movement of the
electrons through DNA. We model the electron as hopping along sites labelled $i$, the
position of a base on the back-bone. We have two electron fields, one on each ribbon, which
we call $\psi$ and $\phi$ (connected to the two sets of coordinates $r$ and $s$ described above).

The action for the electrons, without coupling between the two chains, is
taken to be
\begin{eqnarray}
\mathcal{A}=\int dt\, && \left\{ -\hbar\sum_{i}\psi_{i}^{*}\mathrm{i}\psi_{i,t}-\hbar\sum_{i}\phi_{i}^{*}\mathrm{i}\phi_{i,t}+\right.\nonumber \\
&&
\sum_{i}\left[\hbar\omega2\left|\psi_{i}\right|^{2}-\sum_{j\neq i}J_{i-j}\psi_{i}^{*}\psi_{j}-\frac{1}{2}\chi\left|\psi_{i}\right|^{4}\right]+\nonumber
\\ &&
\left.\sum_{i}\left[\hbar\omega2\left|\phi_{i}\right|^{2}-\sum_{j\neq
i}J_{i-j}\phi_{i}^{*}\phi_{j}-\frac{1}{2}\chi\left|\phi_{i}\right|^{4}\right]\right\}
\quad,\label{eq:action}\end{eqnarray} This can, for instance, be derived
as the semi-classical limit of a tight-binding model, with a
density-squared term added.  Such non-linearities are crucial, since
they limit the amount of charge that can collect at any one site. It
is not obvious that such non-linearities are restricted to the on-site
interaction, and we could also study the question ``how different
would the dynamics of the system be had we added an electronic term of
the density-density correlation form $\sum_{i,j\neq
i}\chi_{ij}\left|\psi_{i}\right|^{2}\left|\psi_{j}\right|^{2}$?''  We
have no answer to this question - and we leave its resolution to some
future work. In this paper we  focus our attention on the study of
the system essentially described by (\ref{eq:action}) together with a
further potential describing the interchain interaction. As we shall
see this system is already sufficiently complicated to exhibit many
interesting properties.

\subsection{The interaction between chains}

Of course, one of the key elements of the problem is the interaction
between the chains (and we limit ourselves to consider only the
interaction between the two nearest sites on the backbones,
i.e. $i=j$).  The most naive choice of such an interaction is to take
\begin{equation}
V_{\text{int}}=\sum_{i}K
\left(\phi_{i}^{*}\psi_{i}+\psi_{i}^{*}\phi_{i}\right)\quad.\label{eq:Vint1}\end{equation}
This form of the interchain interaction term is actually rather
uninteresting since it leads to a simple oscillation of charges from
one strand to the other, while their combination behaves as a single
unit, and we would thus reproduce the usual results of the discrete
nonlinear Schr\"odinger Equation (DNLS)\footnote{For a good
description of properties of this model see e.g.
\cite{Eilbeck2002}.}.

This is due to the fact that Eq.~(\ref{eq:Vint1}) does not take into
account the chirality of the molecules. A possible improvement would
involve making the coupling orientation dependent:
\begin{equation}
V_{\text{int}}=\sum_{i}K(\mathrm{e}^{\mathrm{i}2g\theta_{i}}\phi_{i}^{*}\psi_{i}+
\mathrm{e}^{-\mathrm{i}2g\theta_{i}}\psi_{i}^{*}\phi_{i})\quad,\label{eq:Vint2}
\end{equation}
(where $g$ is yet another parameter).
Here $\theta_i$ is the angle of the $i$ sites on the backbones relative
to some arbitrary direction (see later). This is not 
satisfactory, however, since for a flat molecule the tunnelling would now depend
on its orientation! We can think of two further improvements:

First of all we can subtract the average orientation, \begin{equation}
V_{\text{int}}=\sum_{i}K(\mathrm{e}^{\mathrm{i}2g(\theta_{i}-\langle\theta\rangle)}\phi_{i}^{*}\psi_{i}+\mathrm{e}^{-\mathrm{i}2g(\theta_{i}-\langle\theta\rangle)}\psi_{i}^{*}\phi_{i}\quad,\label{eq:Vint3}\end{equation}
which leads to a rather non-local coupling. This rather artificial
approach will not be further investigated in this work, since it is
not related to the chirality of the DNA molecule.

Secondly, we can invert the transformation of fields obtained from going
from Eq.~(\ref{eq:Vint1}) to Eq.~(\ref{eq:Vint2}), which
corresponds to  a gauge-like
transformation of the naive interaction, Eq.~(\ref{eq:Vint1}).
This implies that we  should make the replacement 
\begin{equation}
\psi_{i}\rightarrow \mathrm{e}^{\mathrm{i}g\theta_{i}}\psi_{i},\quad
\phi_{i}\rightarrow \mathrm{e}^{-\mathrm{i}g\theta_{i}}\phi_{i}
\end{equation}
\emph{everywhere} (i.e., also in all the interaction terms within
each one of the backbones). For a flat molecule this has no
consequence, and the orientation dependence of Eq.~(\ref{eq:Vint2})
disappears, but for a curved molecule, as we shall show below, this
indeed includes effects due to chirality.

Our modification then implies that we can use Eq.~(\ref{eq:Vint1}),
but must replace the non-local interaction along the backbone by 
\begin{equation}
\sum_{i,j\neq i}J_{i-j}\left(\psi_{i}^{*}\psi_{j}+\phi_{i}^{*}\phi_{j}\right)
\rightarrow
\sum_{i,j\neq i}J_{i-j}\left(\mathrm{e}^{\mathrm{i}g(\theta_{j}-\theta_{i})/2}\psi_{i}^{*}\psi_{j}
+\mathrm{e}^{-\mathrm{i}g(\theta_{j}-\theta_{i})/2}\phi_{i}^{*}\phi_{j}\right)\quad.
\end{equation}
This is obviously rotationally invariant. 

\subsection{Continuum limit and chirality}

Let us look at the continuum limit of
\begin{equation}
\sum_{i,j}J_{|i-j|}\mathrm{e}^{\mathrm{i}g(\theta_{j}-\theta_{i})}\psi_{i}^{*}\psi_{j}
\end{equation}
and
\begin{equation}
\sum_{i,j}J_{|i-j|}\mathrm{e}^{-\mathrm{i}g(\theta_{j}-\theta_{i})}\phi_{i}^{*}\phi_{j},
\end{equation}
in order to see that this expression really describes a chiral
molecule.  The second term is simply related to the first by the
replacement $\theta_{i}\rightarrow-\theta_{i}$ and
$\psi\rightarrow\phi$. We thus concentrate on the first term, where we
assume that $J_{|i|}$ falls sufficiently fast so that we need consider
only the nearest-neighbour terms. Calling the contribution of these
terms, divided by $J_{1}$, $k_1$, we find that
\begin{equation}
k_{1}=\sum_{i}\left(\psi_{i}^{*}\psi_{i+1}\mathrm{e}^{\mathrm{i}g(\theta_{i+1}-\theta_{i})}+\text{c.c.}\right).
\end{equation}
Next we assume that $\theta_{i+1}-\theta_{i}$ is small and
we expand  $k_1$ to first order in this quantity. Before doing so, we
introduce some time-saving notation, 
\begin{equation}
(\delta f)_{i}  =  \frac{f_{i+1/2}-f_{i-1/2}}{h},\qquad
\langle f\rangle_{i}  =  \frac{f_{i+1/2}+f_{i-1/2}}{2}.
\end{equation}
Let's now expand and simplify $k_{1}$
\begin{eqnarray}
k_{1} & = & \sum_{i}\left(\psi_{i}^{*}\psi_{i+1}(1+\mathrm{i}g(\delta\theta)_{i+1/2})+\text{c.c.}\right)\nonumber \\
 & = & \frac{1}{2}\sum_{i}\left(\psi_{i}^{*}(\psi_{i+1}+\psi_{i-1}+\text{c.c.}\right)\nonumber \\
 &  & +\frac{g}{2}\sum_{i}\left(\mathrm{i}\psi_{i}^{*}\left(\psi_{i+1}(\delta\theta)_{i+1/2}+\psi_{i-1}(\delta\theta)_{i-1/2}\right)+\text{c.c.}\right)\nonumber \\
 & = & \frac{h^{2}}{2}\sum_{i}\left(\psi_{i}^{*}(\delta^{2}\psi)_{i}+\text{c.c.}\right)+\sum_{i}\left|\psi_{i}\right|^{2}\nonumber \\
 &  & +h\frac{g}{2}\sum_{i}\left(\mathrm{i}\psi_{i}^{*}\left(\langle\psi\rangle_{i+1/2}(\delta\theta)_{i+1/2}-\langle\psi\rangle_{i-1/2}(\delta\theta)_{i-1/2}\right)+\text{c.c.}\right)\nonumber \\
 & = & \frac{h^{2}}{2}\sum_{i}\left(\psi_{i}^{*}(\delta^{2}\psi)_{i}+\text{c.c.}\right)+\sum_{i}\left|\psi_{i}\right|^{2}\nonumber \\
 &  & +h^{2}\sum_{i}\left(\psi_{i}^{*}\biggl(\delta\bigl[\langle\psi\rangle\bigl(i\frac{g}{2}\delta\theta\bigr)\bigr]\biggr)_{i}+\text{c.c.}\right).\end{eqnarray}
If we combine the ``onsite'' term $\left|\psi_{i}\right|^{2}$ in a similar
term already present in the model,
we find that the remaining terms become
\begin{eqnarray}
k_{1}/h & = & \frac{1}{2}\int dx\,
\left(\psi^{*}\partial_{x}(\partial_{x}+\mathrm{i}gA)\psi+
\psi\partial_{x}(\partial_{x}-\mathrm{i}gA)\psi^{*}\right)\nonumber
\\ & \approx & -\int dx\,\left((D_{x}\psi)^{*}D_{x}\psi\right).
\end{eqnarray} 
Here
$A=\partial_{x}\theta$, $D_{x}=\partial_{x}+\mathrm{i}gA$. In the last step we
have added a second order term in $\theta$ (neglected in our initial
evaluation).  Thus we end up with a covariant derivative. The
other chain ($\phi$) contains the complex conjugate covariant derivative, 
\begin{equation}
k_{1}^{\phi}\approx-\int dx\,\left((D_{x}^{*}\phi)^{*}D_{x}^{*}\phi\right).
\end{equation}
Thus we see that with this structure of the coupling to a `magnetic
flux' caused by the warping of the molecule through each plaquette, we
have indeed constructed a chiral model of electrons moving through a
molecule.

\subsection{Solitons in a curved background}
First we shall study the discrete solitons (breathers) of the model,
i.e., solutions of the form $\psi_i=\mathrm{e}^{\mathrm{i}\Omega t} f_{-1,i}$,
$\phi_i=\mathrm{e}^{\mathrm{i}\Omega t} f_{1,i}$. We assume that the distance between the
molecules is specified by the parameters $d = 2.39\text{ nm}$, $\delta
z = 3.38/10=0.338\text{ nm}$, $\delta\theta = 2\pi/10=0.628$.

This leads to the non-linear eigenvalue problem
\begin{equation}
 2 \hbar \omega f_{k,i}-\sum_{j\neq i} 
\mathrm{e}^{\mathrm{i}g\lambda (\theta_{j}-\theta_{i})/2}J_{i-j}f_{\lambda,j}
-\chi |f_{\lambda,i}|^2f_{\lambda,i} +K f_{\bar \lambda, i}=
\hbar \Omega f_{\lambda,i}\quad,
\end{equation}
where the term containing $\bar\lambda=-\lambda$ denotes the coupling
to the other chain. One of the ways this problem can be solved is 
through a
constrained minimization, with $\hbar \Omega$ being the Lagrange multiplier
imposing charge normalisation.

We have studied the nature of such solutions, for $\hbar\omega=1$,
$J_{i-j}=J_0\mathrm{e}^{-\beta d_{ij}}$, $\chi=0.1$ and $\kappa=1$
We have found that the coupling between the two chains leads 
to solitons that are distributed
evenly over the two backbones, but where the phase varies.

\begin{figure}
\begin{center}\includegraphics[width=6cm,  keepaspectratio,clip]{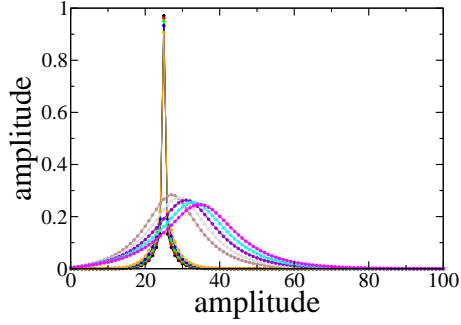}\end{center}
\caption{\label{fig:DNLS}
The profile of the DNLS breather for various values
of the coupling constant $J_0$, starting from $J_0=0.22$ and a solution localized at $i=25$, increasing $J_0$ in syteps of $0.02$. The transition occurs between $J_0=0.3$ and $J_0=0.32$. Wider curves represent larger values of $J_0$.
}
\end{figure}
\begin{figure}
\begin{center}\includegraphics[width=6cm,  keepaspectratio,clip]{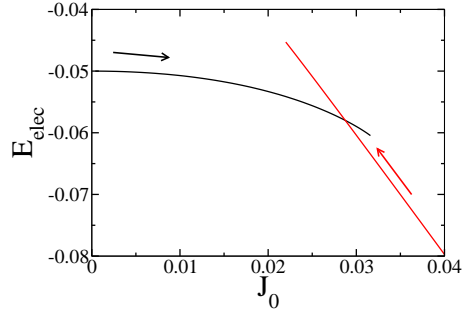}\end{center}
\caption{\label{fig:PT}The transition between solitonic and delocalised solutions as a function 
of the coupling constant $J_0$.}
\end{figure}

If we ignore the coupling between the two strands of the double helix
for a while, and analyse the motion of a single electron along an
equilibrium chain, we are essentially solving the DNLS, and the
interesting solutions are breathers of the form $\phi_i(t)=\mathrm{e}^{\mathrm{i}\omega
t} f_i$ \cite{Eilbeck2002}. These can be found quite easily for the
current model by a simple steepest descent method (minimizing the
electronic energy). 
In figure~\ref{fig:DNLS} we present the profiles of the DNLS breather
for various values of the coupling constant $J_0$.
In our simulations we have found a sharp transition from a localised to delocalised
solution for values of $J_0$ about $0.0282$.
As can be seen from figure~\ref{fig:PT} the transition appears  to be of
first order, and the solutions are metastable beyond the transition.

\begin{figure}
\begin{center}\includegraphics[width=6cm,  keepaspectratio,clip]{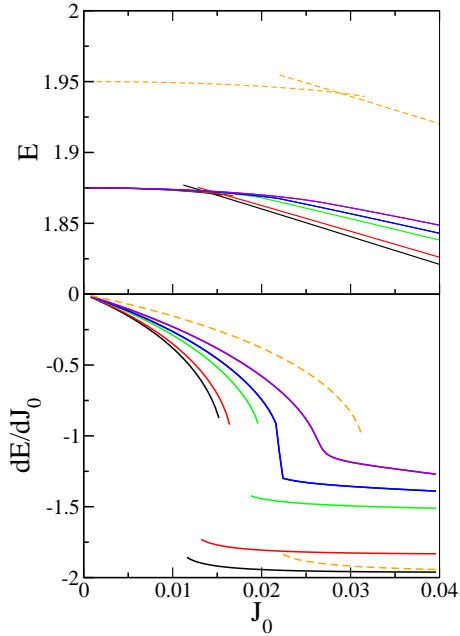}\end{center}
\caption{\label{fig:beta1E}The energy (mass) of the breather for two coupled chains, as a function
of the coupling constant $J_0$, and $\beta=1 \text{nm}^{-1}$. From
bottom to top, the curves correspond to: black $g=0$, red $g=0.5$,
green $g=1$, blue $g=1.2$ and violet $g=1.5$.  The dashed (orange)
curves show the single chain solution from above.  In each case we
have performed our calculations for both the increasing and decreasing
$J_0$.}
\end{figure}
\begin{figure}
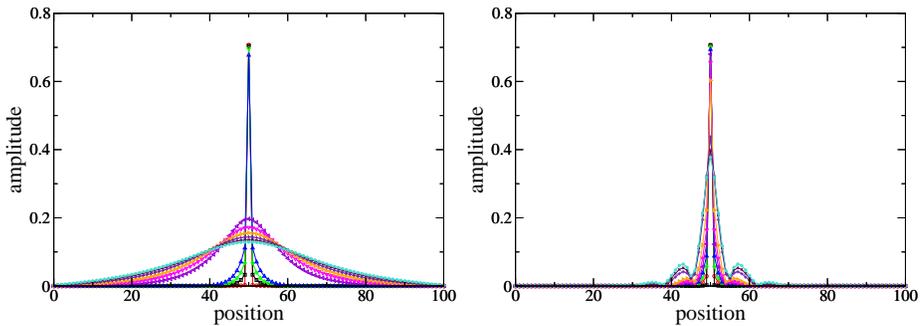

\begin{center}\includegraphics[width=6cm,  keepaspectratio,clip]{beta1g0.eps}
\includegraphics[width=6cm,  keepaspectratio,clip]{beta1g1p5.eps}\end{center}
\caption{\label{fig:beta1prof}The profile of the DNLS breathers as a function of the coupling constant 
of the coupling constant $J_0$, and  $\beta=1 \text{nm}^{-1}$. The left figure shows $g=0$, the right one $g=1.5$. The sharpest curves occur for the smallest value of
$J_0$, with a sequential broadening on increase.}
\end{figure}

\begin{figure}
\begin{center}\includegraphics[width=6cm,  keepaspectratio,clip]{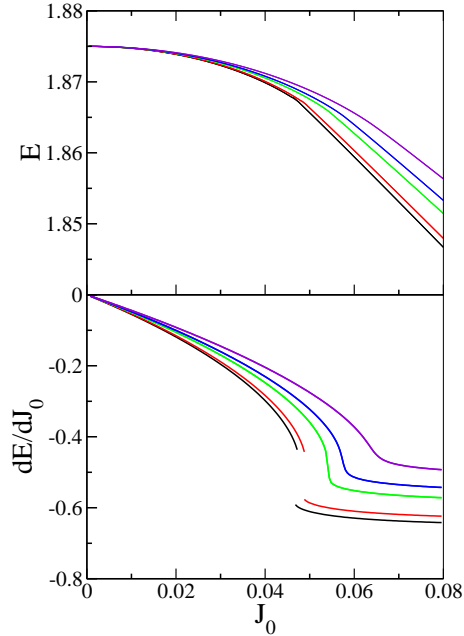}\end{center}
\caption{\label{fig:beta2E}The energy (mass) of the breather for two coupled chains, as a function
of the coupling constant $J_0$, and $\beta=2 \text{nm}^{-1}$. From
bottom to top, the curves correspond to: black $g=0$, red $g=0.5$,
green $g=1$, blue $g=1.2$ and violet $g=1.5$.  In each case we
have performed our calculations for both the increasing and decreasing
$J_0$.}
\end{figure}

\begin{figure}
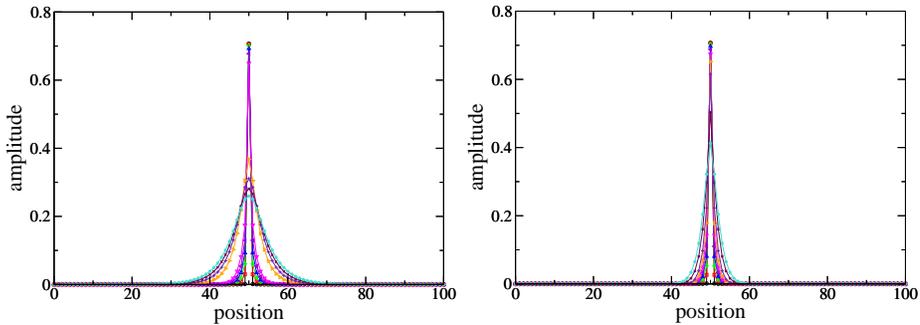

\begin{center}\includegraphics[width=6cm,  keepaspectratio,clip]{beta2g0.eps}
\includegraphics[width=6cm,  keepaspectratio,clip]{beta2g1p5.eps}\end{center}
\caption{\label{fig:beta2prof}The profile of the DNLS breathers as
 a function of the coupling constant  $J_0$, and  $\beta=2 \text{nm}^{-1}$. The left figure shows $g=0$, the right one $g=1.5$. The sharpest curves occur for the smallest value of
$J_0$, with a sequential broadening as $J_0$ increases.}
\end{figure}

When we increase the coupling we find that it is instructive to look
in more detail at the energetics of the problem, as shown in
figure~\ref{fig:beta1E}, where $\beta=1 \text{nm}^{-1}$ and various
values of $g$ have been used.  We still see the phase trasition, but
note that there is a shift in the position, and that it seems to
change from first to second or higher order with increasing chiral
charge $g$. As can be seen in figure \ref{fig:beta1prof}, there is a change in the
nature of the solitonic positions as well, and we are no longer able
to find fully delocalised ones. If we increase $\beta$ the transition
seems to weaken even more, and the solitons seem to sharpen; see
figures \ref{fig:beta2E} and \ref{fig:beta2prof}.


\begin{figure}
\begin{center}\includegraphics[width=12cm,  keepaspectratio,clip]{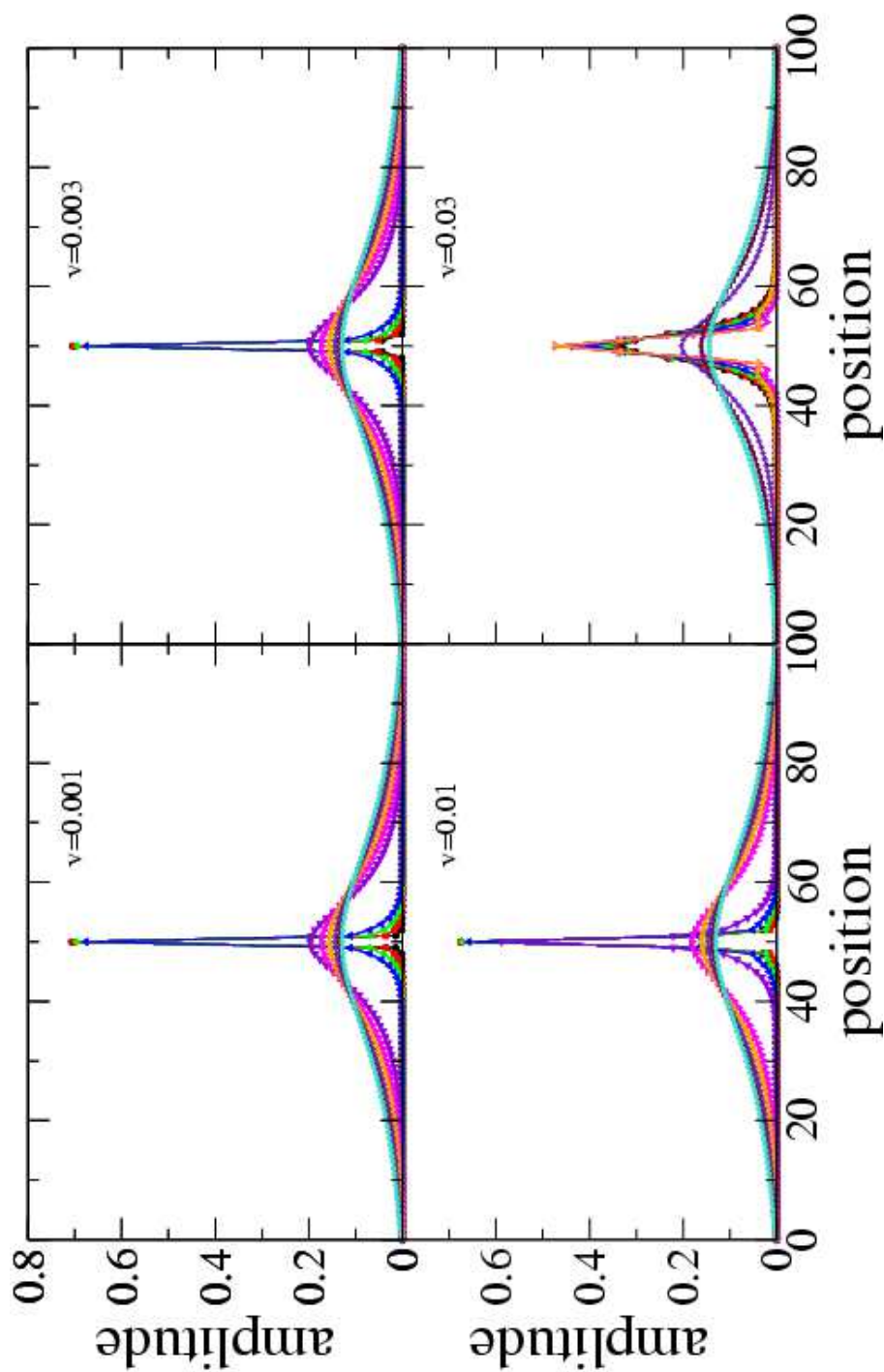}\end{center}
\caption{\label{fig:beta2g0mv}The shape of the moving breather for 
two coupled chains, as a function
of the coupling constant $J_0$, $g=0$ and $\beta=2 \text{nm}^{-1}$. The
sharpest curves occur for the smallest value of $J_0$. }
\end{figure}

\begin{figure}
\begin{center}\includegraphics[width=12cm,  keepaspectratio,clip]{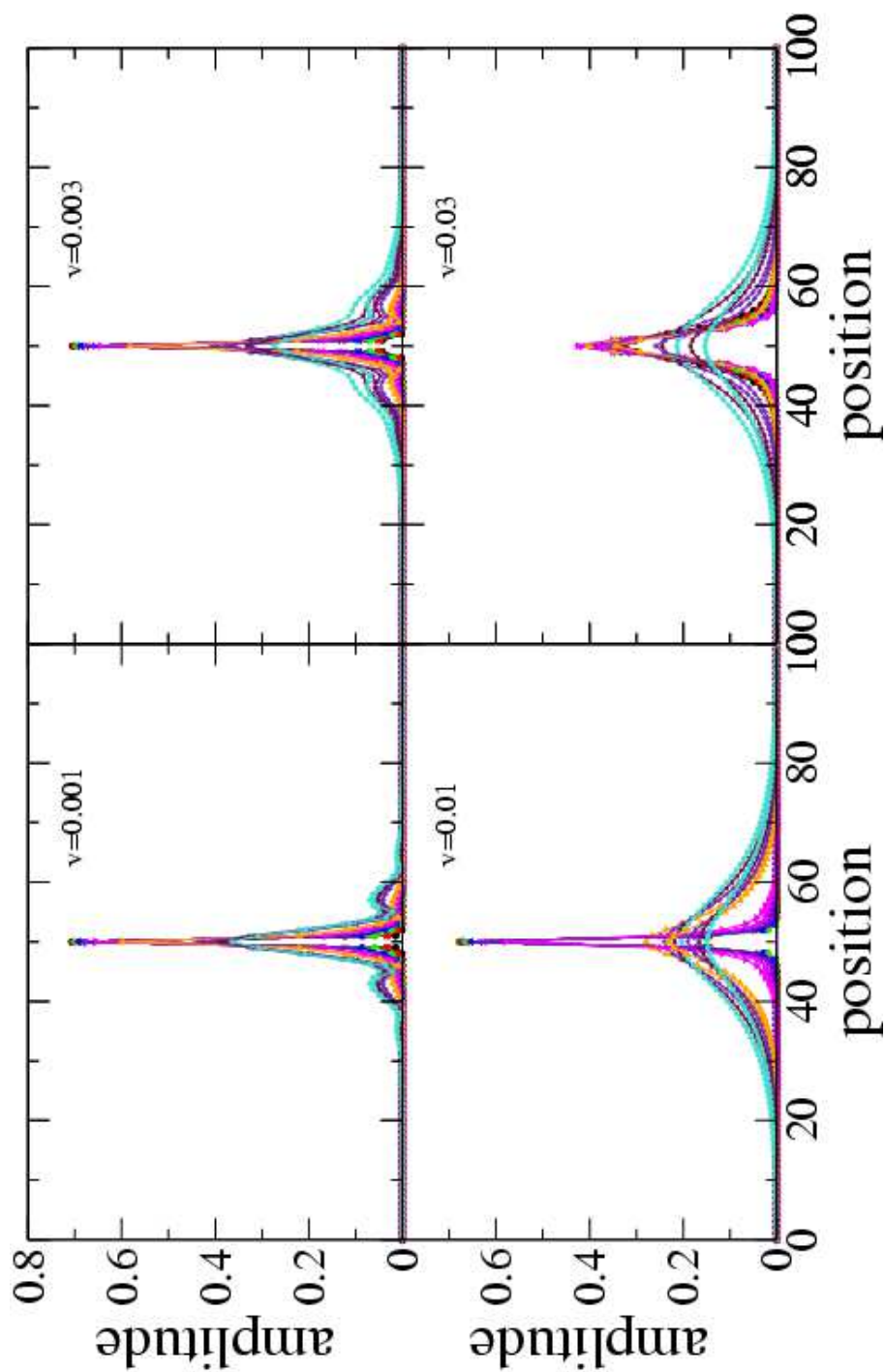}\end{center}
\caption{\label{fig:beta2g1p5mv}
The shape of the moving breather for two coupled chains, as a function
of the coupling constant $J_0$, $g=1.5$ and $\beta=2
\text{nm}^{-1}$. The sharpest curves occur for the smallest value of
$J_0$ broadening as $J_0$ increases.}
\end{figure}

If we wish to study the mobility of the solitons, moving breathers, etc
 the standard technique \cite{Eilbeck2002}
 in the
field suggests the use of the smooth-profile (continuum) 
approximation, where we have added a term of the form 
\begin{equation}
\mathrm{i}v (f_{\lambda,i+1}-f_{\lambda,i-1})/2
\end{equation}
to the left hand-side of the eigenvalue problem. 
Results of such calculations are given in figures \ref{fig:beta2g0mv} and
\ref{fig:beta2g1p5mv}. Here the phase
transition probably plays an important role as for small $J_0$ we have a
very localised solution which strongly violates the continuum
approximation.  In this case we are unable to find a moving
soliton. For larger values of $J_0$ we do find a smooth profile
solution, which suggests that we have found the moving soliton. This
we expect to have an important impact on the charge transport properties
of the system. Generally, the
effect of the chirality of the molecule seems  to sharpen
the profile thus reducing the mobility!

\subsection{The molecule}

The model for the molecule can be specified in its most general form
by solving a number of constraints, which we shall try to parametrise
in as simple a way as possible.  Looking at the figure, it seems easy
to work with the following representation: (we use $\vec{r}_{i}$ to
denote the position where the $i$th base pair is attached to one of
the chains, $\vec{s}_{i}$ for the other one) 
\begin{equation}
\vec{r}_{i}=z_{i}\vec{e}_{z}+\frac{1}{2}d\vec{u}_{i},\quad\vec{s}_{i}=z_{i}\vec{e}_{z}-\frac{1}{2}d\vec{u}_{i},
\end{equation}
with $\vec{u}$ a 2D unit vector in the $xy$ plane. 
Hence it is natural to relate the angles $\theta_i$ introduced in (3)
to the orientation of $\vec{u}_i$ in the $xy$ plane. Our representation
makes it possible to allow for stretching of the $H$ bonds by adding
a potential for $d$ (as in the Peyrard-Bishop model). We shall use
a molecular potential to implement an approximate constraint on the
spacing in the strands of DNA, but it is not so straightforward to
study the effect of bending of the double helix in this model,
which could be quite important \cite{Levitt83}.

The kinetic energy is relatively easy to evaluate 
\begin{eqnarray}
\frac{1}{2}M\sum_{i}\vec{r}_{i,t}^{2}+\vec{s}_{i,t}^{2} & = & \frac{1}{2}M\sum_{i}\left(z_{i}\vec{e}_{z}+\frac{1}{2}d\vec{u}_{i}\right)_{t}^{2}+\left(z_{i}\vec{e}_{z}-\frac{1}{2}d\vec{u}_{i}\right)_{t}^{2}\nonumber \\
 & = &
 M\sum_{i}\left(z_{i,t}^{2}+\frac{1}{4}d^{2}\vec{u}_{i,t}^{2}\right)\quad.
\end{eqnarray}
We can easily generalise this to a dynamical $d$; in this case the
length of each rung becomes $d_{i}(t)$, and the additional terms in
the kinetic energy would be of the form
$\frac{1}{4}d_{i,t}^{2}$. However, we expect the effects of the
variation of $d$ to be small, so in this work we will keep $d$ fixed.

We shall now describe a set of simple potentials, all corresponding to
properties of the bonds or the angles between bonds, that can be used
to describe equilibrium DNA.

If we insist on a molecule where the equilibrium shape is bent, such as
DNA, we  must impose an equilibrium angle between the bonds. One 
such potential would be
\begin{eqnarray}
U_{B} & \propto & \sum_{i}\left[\left((\vec{r}_{i}-\vec{r}_{i-1})\times(\vec{r}_{i+1}-\vec{r}_{i})\right)^{2}-\alpha_{s}^{2}\right]^{2}\nonumber\\&&
+\left[\left((\vec{s}_{i}-\vec{s}_{i-1})\times(\vec{s}_{i+1}-\vec{s}_{i})\right)^{2}-\alpha_{s}^{2}\right]^{2}
\quad.
\end{eqnarray}
Actually, this is too simple. In turning the outer product into a scalar, we
have now allowed  mixing of the left- and
right-handed turns in the same molecule.
Since DNA is a chiral molecule,  we would like to add
a term that prefers left-handed over right-handed twists. This is most
easy to construct the continuum limit, where this translates into
the requirement that the vector tangential to one of the strands  points
both upwards and into the plane, or in the opposite direction (downwards and out of
the plane).  Suppose that the point on the strand is specified by a
vector $\vec{\delta}$ from the back bone (i.e., $\vec{\delta}$ lies in
the $xy$ plane), we write $\vec{\delta}=\delta\vec{e}$ (in the
notation used above, $\vec{e}=\pm\vec{u}$, depending on which strand
we are on), with
\begin{equation}
\vec{e}=(\cos\theta,\sin\theta,0).\end{equation}
We can then write for the tangent vector
\begin{equation}
\vec{t}=a(-\sin\theta,\cos\theta,0)+b(0,0,1),\end{equation}
and, to introduce chirality, we wish to add a bias
towards a particular value of $ab$  in our model. We can extract $a$ by calculating
$\vec{e}_{z}\cdot(\vec{e}\times\vec{t)}$, and $b=\vec{e_{z}}\cdot\vec{t}$.
Replacing $\vec{t}$ by the discrete analogue $\vec{r}_{i+1}-\vec{r}_{i}$, we find that
\begin{eqnarray}
a_{r} & \approx &
\vec{e}_{z}\cdot\left[\frac{1}{2}(\vec{u}_{i}+\vec{u}_{i+1})\times\left((z_{i+1}-z_{i})\vec{e}_{z}-\frac{d}{2}(\vec{u}_{i+1}-\vec{u}_{i}\right)\right]\nonumber
\\ & = &
\frac{d}{2}\vec{e}_{z}\cdot(\vec{u}_{i+1}\times\vec{u}_{i})\quad,\\
b_{r} & \approx &
\vec{e}_{z}\cdot\left((z_{i+1}-z_{i})\vec{e}_{z}-\frac{d}{2}(\vec{u}_{i+1}-\vec{u}_{i})\right)\nonumber
\\ & = & z_{i+1}-z_{i}\quad.\end{eqnarray} Here, due to the symmetry,
$a_{s}=a_{r}$ and $b_{s}=b_{r}$. A potential that achieves our aim is
of the form\begin{equation}
U_{B}=\sum_{i}g_{\mu}\bigl((z_{i+1}-z_{i})\vec{e}_{z}\cdot(\vec{u}_{i+1}\times\vec{u}_{i})\bigr)\quad,\end{equation}
with $g$ a function that has a minimum at $\mu$, the
chosen average parameter, and increases quickly as we move away from
equilibrium. An example of such a function which in fact we have used
in our simulations, is
\begin{equation}
g(x)=b(x-\mu)^{2}.\label{eq:bending}\end{equation}

We will also assume that each of the back-bones (the strands) can
stretch; this will be built into our model by introducing a stretching
potential of the bonds of the form (note that stretching for the other
strand is constrained by the symmetry imposed by the straightness of
our molecule)
\begin{eqnarray} U_{s} & = &
\sum_{i}f_{l}(|\vec{r}_{i}-\vec{r}_{i+1}|^{2})\nonumber \\ & = &
\sum_{i}f_{l^{2}}((z_{i+1}-z_{i})^{2}+\frac{d^{2}}{2}(1-\vec{u}_{i+1}\cdot\vec{u}_{i}))\quad,\end{eqnarray}
where $f$ is has a minimum at the average distance $l^{2}$, and
increases quickly away from the minimum. As $f$
describes a stretching potential we can either use a harmonic
potential, $f(x)=a(x-l^{2})^{2}$, or a Morse potential, $f(x)=a(\mathrm{e}^{-(x-l^{2})^{1/2}}-1)^{2}$.
 In our simulations we
have used the simple quadratic form
\begin{equation}
f(x)=a(x-l^{2})^{2}\label{eq:stretching}\end{equation}

\subsection{A small simplification}

Before we proceed further we note that we can simplify the problem slightly by using a 1D angle representation
for $\vec{u}$, \begin{equation}
\vec{u}_{i}=(\cos\theta_{i},\sin\theta_{i},0)\quad.\end{equation}
 We then find\begin{equation}
\vec{u}_{i,t}^{2}=\dot{\theta}_{i}^{2}\quad,\end{equation}
 and \begin{equation}
U_{B}=\sum_{i}g_{\mu}((z_{i+1}-z_{i})\sin(\theta_{i+1}-\theta_{i}))\quad,\end{equation}
 as well as\begin{eqnarray}
U_{s} & = & \sum_{i}f_{l^{2}}((z_{i+1}-z_{i})^{2}+\frac{d^{2}}{2}(1-\cos(\theta_{i+1}-\theta_{i})))\quad.\end{eqnarray}

Next we determine the equations of motion for all the fields of our
system. They are given by:

\begin{equation}
-\hbar \mathrm{i}\psi_{i,t}=-2\hbar\omega\psi_{i}+\sum_{j\neq i}J_{i-j}\psi_{j}+\frac{1}{2}\chi\psi_{i}^{2}\psi_{i}^{*}+K\phi_{i}\quad,\end{equation}

\begin{eqnarray}
\frac{d^{2}}{2}M\ddot{\theta}_{i} & = &
 +g'((z_{i+1}-z_{i})\sin(\theta_{i+1}-\theta_{i}))(z_{i+1}-z_{i})\cos(\theta_{i+1}-\theta_{i})\nonumber
 \\ & &
 -g'((z_{i}-z_{i-1})\sin(\theta_{i}-\theta_{i-1}))(z_{i}-z_{i-1})\cos(\theta_{i}-\theta_{i-1})\nonumber
 \\ & &
 +f'((z_{i+1}-z_{i})^{2}+\frac{d^{2}}{2}(1-\cos(\theta_{i+1}-\theta_{i})))\frac{d^{2}}{2}\sin(\theta_{i+1}-\theta_{i}))\nonumber
 \\ & &
 -f'((z_{i-1}-z_{i})^{2}+\frac{d^{2}}{2}(1-\cos(\theta_{i}-\theta_{i-1})))\frac{d^{2}}{2}\sin(\theta_{i}-\theta_{i-1}))-\alpha
 \dot \theta_i\nonumber \\ & & +\text{terms depending on $J$}
\label{eq:thetaeom}
\end{eqnarray} 
and 
\begin{eqnarray} 2M\ddot{z}_{i} & = &
 +g'((z_{i+1}-z_{i})\sin(\theta_{i+1}-\theta_{i}))\sin(\theta_{i+1}-\theta_{i})\nonumber
 \\ & &
 -g'((z_{i}-z_{i-1})\sin(\theta_{i}-\theta_{i-1}))\sin(\theta_{i}-\theta_{i-1})\nonumber
 \\ & &
 +f'((z_{i+1}-z_{i})^{2}+\frac{d^{2}}{2}(1-\cos(\theta_{i+1}-\theta_{i})))2(z_{i+1}-z_{i})\nonumber
 \\ & &
 -f'((z_{i}-z_{i-1})^{2}+\frac{d^{2}}{2}(1-\cos(\theta_{i}-\theta_{i-1})))2(z_{i}-z_{i-1})-\alpha\dot
 z_i\nonumber \\ & & +\text{terms depending on
 $J$}\quad.
\label{eq:zeom}
\end{eqnarray} 
Please note that we have added a damping term to both the $z$ and
$\theta$ equations (the term proportional to $\alpha$). This helps us
to lower the energy in our system and to determine its lowest energy
states which will be solitonic if the system possesses such solutions.

The only remaining question is the choice of $g$. A sensible choice
seems to be: $g_{\mu}(x)=b(x-\mu)^{2}$ and this is the expression we
used in our simulations.

\subsection{$J$ dependence}

Next we add the $J$ dependent terms which will control electron's
mobility. Although, in principle, $J$ can depend on many variables
two obvious dependences first spring to mind:

\begin{enumerate}
\item The tunnelling rate is a function of the shortest distance between
two spatial points, i.e., $r_{ij}=|\vec{r}_{i}-\vec{r}_{j}|$, or 
\item The tunnelling rate depends on the distance along the DNA backbone
(the sucrose molecules), $r_{ij}=\sum_{k=i}^{j-1}|\vec{r}_{i}-\vec{r}_{i+1}|$. 
\end{enumerate}
The first case is obviously easier to implement,but shall not be used in this
work,

The second case is more complicated. Nonetheless; we can write $J_{ij}=h(R_{ij})$
and find that the additional terms in the EoM for $\theta$, Eq.~(\ref{eq:thetaeom})
are of the form
\begin{eqnarray}
&&-\frac{d^{2}\sin(\theta_{i+1}-\theta_{i})}{r_{i+1,i}}\sum_{k\leq
i,l>i}h'(R_{kl})\psi_{k}^{*}\psi_{l}+(k\leftrightarrow l)\nonumber\\&&
-\frac{d^{2}\sin(\theta_{i-1}-\theta_{i})}{r_{i-1,i}}\sum_{k<i,l\geq
i}h'(R_{kl})\psi_{k}^{*}\psi_{l}+(k\leftrightarrow l)
\end{eqnarray}
and for $z$ (Eq.~(\ref{eq:zeom}))
\begin{eqnarray}
&&-\frac{d^{2}z_{i}(z_{i}-z_{i+1})}{r_{i+1,i}}\sum_{k\leq
i,l>i}h'(R_{kl})\psi_{k}^{*}\psi_{l}+(k\leftrightarrow l)\nonumber\\&&
-\frac{d^{2}z_{i}(z_{i}-z_{i-1})}{r_{i-1,i}}\sum_{k<i,l\geq
i}h'(R_{kl})\psi_{k}^{*}\psi_{l}+(k\leftrightarrow l)\quad.
\end{eqnarray}

\subsection{Choice of parameters}

Next we attempt to determine the realistic (i.e. of some relevance in
biophysics) range of our parameters. First we look at the static
solutions of the {}``molecular'' part of the energy, and minimise both
the bending and the stretching potentials; we find that
\begin{eqnarray}
(z_{i+1}-z_{i})\vec{e}_{z}\cdot(\vec{u}_{i}\times\vec{u}_{i+1}) & = &
\mu\quad,\\
(z_{i+1}-z_{i})^{2}-\frac{d^{2}}{2}\vec{u}_{i+1}\cdot\vec{u}_{i} & = &
l^{2}-\frac{d^{2}}{2}.\end{eqnarray} 
Assuming uniformity we get
\begin{eqnarray}
\delta z\sin\delta\theta & = & \mu\quad,\\
\delta z^{2}+d^{2}\sin^{2}(\delta\theta/2) & = & l^{2}\quad,
\end{eqnarray}
where $\delta z=(z_{i+1}-z_{i}$), $\delta\theta$ is the angle between
two subsequent $\vec{u}$'s. Thus it is simple to find a value for
$\mu$ and $l$ given $\delta z$, $d$ and $\delta\theta$, which we take
from the studies of the real DNA. The standard values are
\cite{CallandineD97}
\begin{eqnarray}
d & = & 2.39\text{ nm}\quad,\\
\delta z & = & 3.38/10=0.338\text{ nm}\quad,\\
\delta\theta & = & 2\pi/10=0.628\quad.\end{eqnarray}
This gives us 
\begin{equation}
\mu=0.20\text{ nm},\quad l=0.81\text{ \text{nm}}\quad.\end{equation}
If we look at all the minima for these values of $\mu$, $l$ and $d$ we
find that we have a second solution 
\begin{equation}
\delta z=0.746\text{ nm},\quad\delta\theta=0.858\quad;\end{equation}
this indicates an obvious weakness of our model---it will lead to
coexisting shapes!

\section{Numerical Studies}

We have performed several numerical studies of our system. We have
taken sensible parameters; i.e. those that appeared not to be completely
inconsistent with the natural physical constraints but, at the same
time, are quite useful to gain some understanding of the nature of the
solutions.  We have taken $\hbar=M=1$, $\omega=1$ (\ref{eq:action}),
$a=6,l^2=0.7$ (\ref{eq:stretching}), $b=6, \mu=0.2$
(\ref{eq:bending}), $K=0.1$ (\ref{eq:Vint2}).  The damping for the
molecular motion was set at $\alpha=0.02$.  We also varied the value
of $J$ from 0.01 to 10.

We have found that for small values of $J$ the system behaved in a
sensible way - i.e., the molecules were quite rigid and the electron
field modified their behaviour very little. The electron field was
solitonic in nature and it affected the molecule primarily in the
places where it was nonzero.  For larger values of $J$---i.e.,
$J>0.035$ the electron field dispersed and its effects were more
pronounced. This can be most easily seen by looking at the figures
which we include. To obtain them we used a 101 {}``rung'' model of the
twisted DNA-ladder, i.e., with 10 complete turns. We note that the
transition point from the localised to the delocalised regime appears
to have moved a bit as compared to the DNLS, see below.

\begin{figure}
\begin{center}\includegraphics[width=7cm,keepaspectratio,clip]{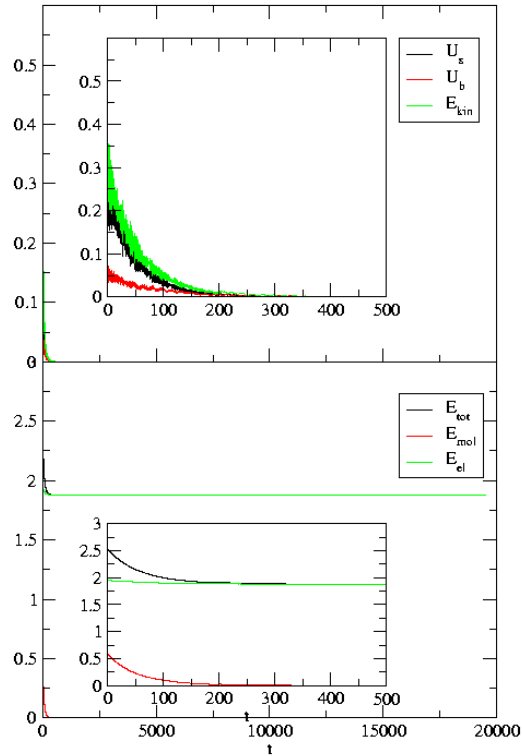}\end{center}

\caption{\label{fig:allp02} Energetics of the relaxation of a 101-rung chain
for $J=0.02$. The upper panel with inset shows the decrease of the
mechanical energy, and the lower panel the same for the total and
electronic energy}
\end{figure}

\begin{figure}
\begin{center}\includegraphics[width=10cm,keepaspectratio,clip]{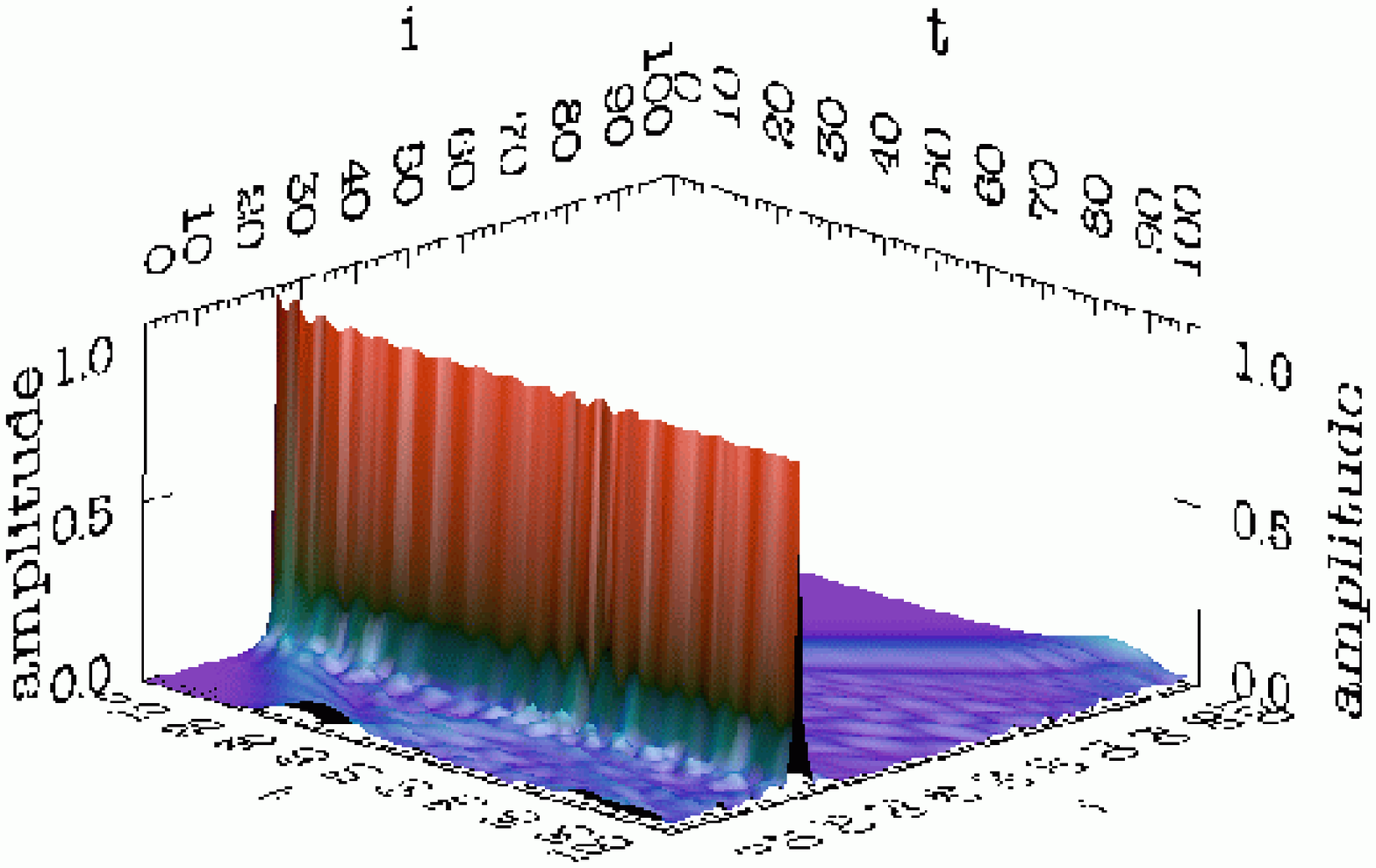} \includegraphics[width=10cm,keepaspectratio,clip]{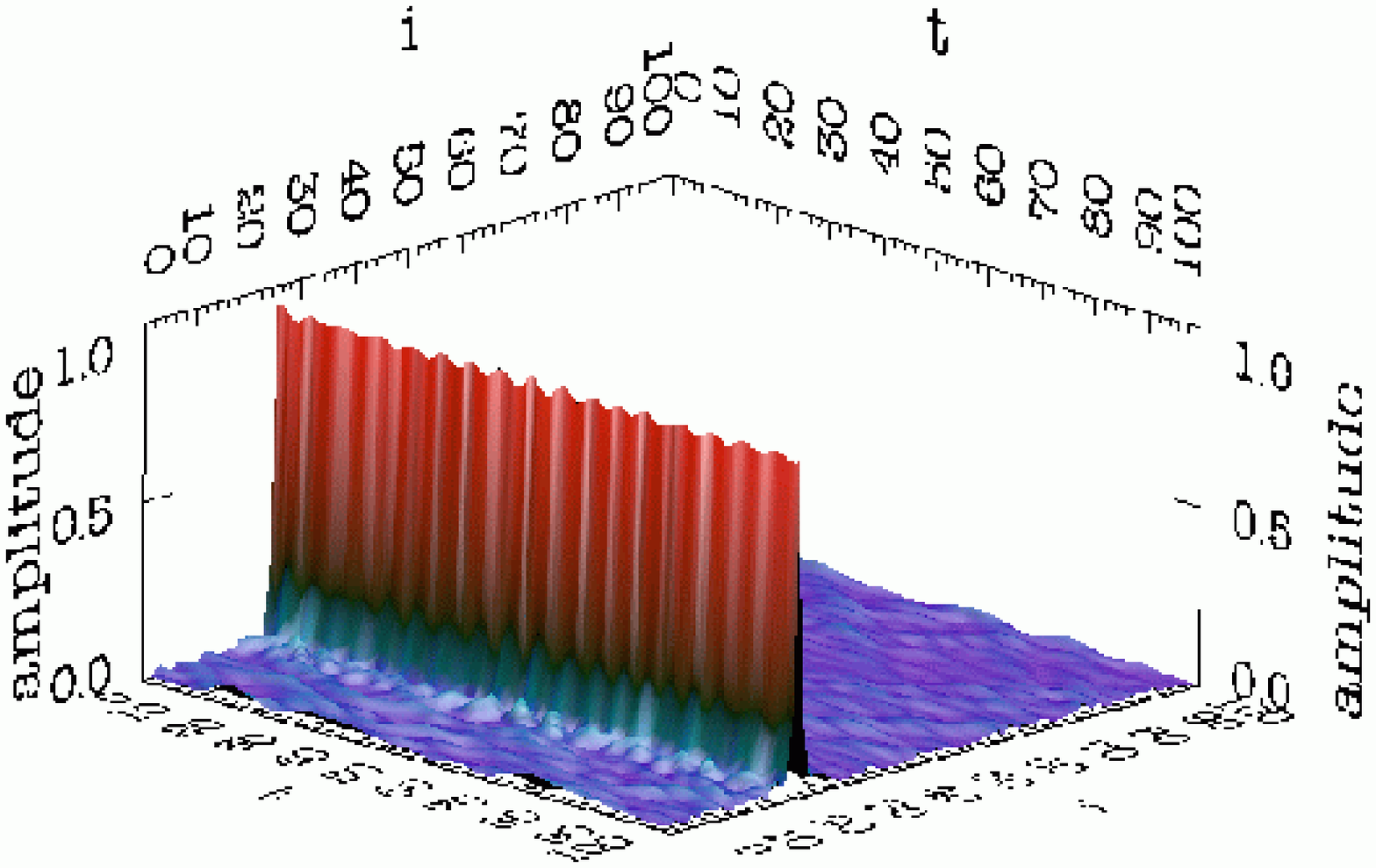}\end{center}

\caption{\label{fig:normp02} Time evolution of the occupation as a function
of the position along the backbone (sum over both) for $J=0.02$.  $t$
gives a time slice number (100 slices used for each figure).  The top
figure shows the early time evolution, the lower one the late time
steady state breather solution. A dynamical representation of the
evolution is available at http://walet.phy.umist.ac.uk/DNA/Jp02.php.}
\end{figure}

In our first simulations we looked at $J=0.02$. As can be seen from
figure~\ref{fig:allp02}, in this case, the kinetic energy decayed
rapidly, and so did the potential, leaving us with a breather soliton,
see figure~\ref{fig:normp02}.

In this figure we show that the amplitude (square-root of the
probability) of the electron field at one rung (i.e. summed over the
two sites connected by that rung) performs semi-regular
oscillations. The simulation was started out with all probability on
one site of one backbone. We note also that some probability starts
propagating outwards. This is then reflected from the boundaries,
returns to the soliton and makes the breather's oscillations somewhat
less periodic resulting in a small sea of background oscillations.

\begin{figure}
\begin{center}\includegraphics[width=10cm,keepaspectratio,clip]{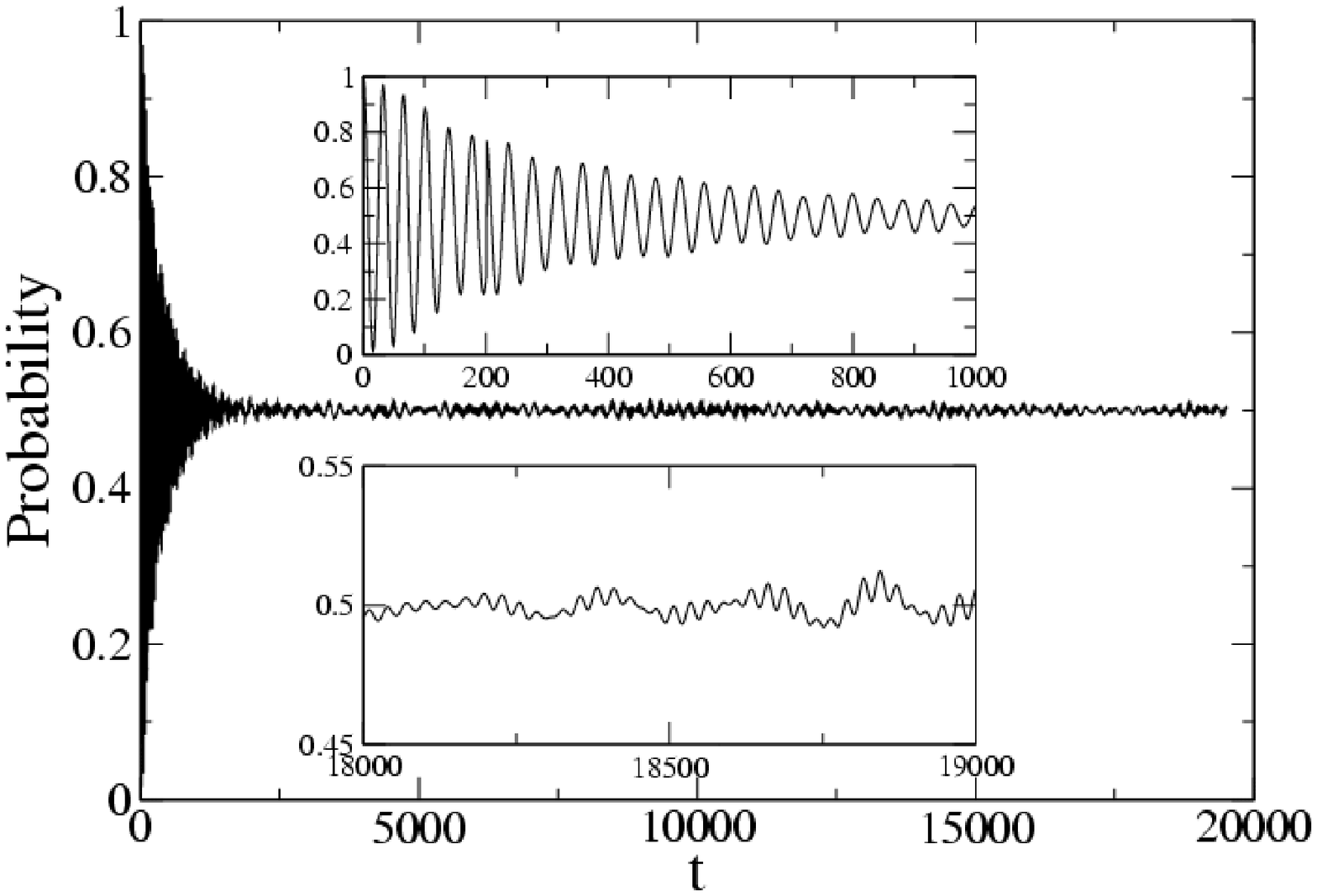}\end{center}

\caption{\label{fig:normbackbonep02} Time evolution of the total occupation
of one of the backbones as a function of time for $J=0.02$. The insets
show both early and late oscillations of this quantity. }
\end{figure}

Some aspects of the periodicity can also be seen in
figure~\ref{fig:normbackbonep02} where we have plotted the time
dependence of the total probability on one of the two backbones
(i.e. summed over all 101 sites on that backbone). We observe a fast
decay to 0.5, and then small oscillations around this value, related
to the periodicity of the breather.

\begin{figure}
\begin{center}\includegraphics[width=10cm,keepaspectratio,clip]{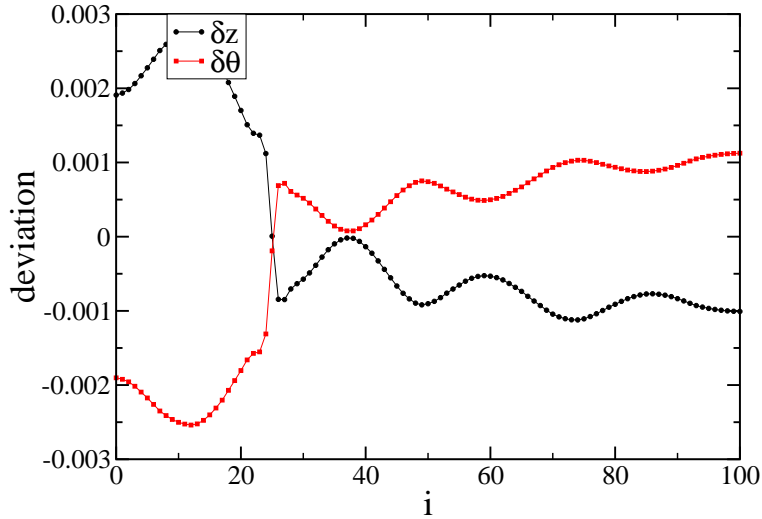}\end{center}

\caption{\label{fig:xzp02} Deviation of $\theta$ (red squares) and $z$ (black circles)
from their equilibrium values for the final simulation point,
$t=19318$, as a function of the lattice point labelled by $i$. The
coupling parameter $J=0.02$. }
\end{figure}

We note that the presence of the soliton leads to a pronounced
deviation of $\theta$ and $z$ from their equilibrium values, with a
sharp jump at the position of the soliton (figure~\ref{fig:xzp02}).
There seems to be a stronger correlation between these two quantities
than one would have naively expected, but the overall picture shows
that there is a deformation and relaxation of the double helix.

Next we have studied a transitional case, namely, $J=0.03$.

\begin{figure}
\begin{center}\includegraphics[width=7cm,keepaspectratio,clip]{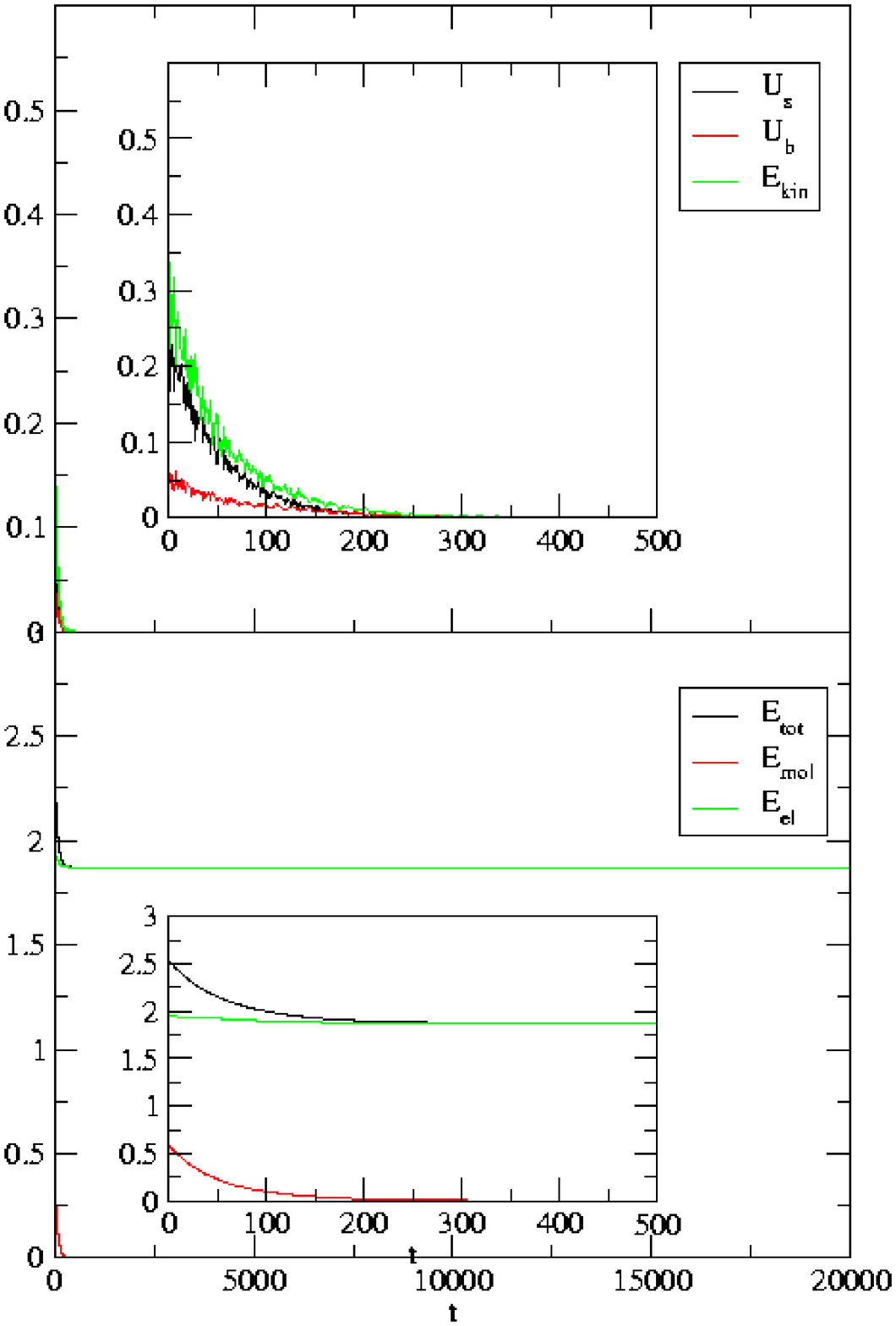}\end{center}

\caption{\label{fig:allp03} Energetics of the relaxation of a 101-rung chain
for $J=0.03$. The upper panel with inset shows the decrease of the
mechanical energy, the lower panel the decrease of the total and the
electronic energy.}
\end{figure}

Once again the relaxation to the equilibrium has been quick and
uneventful, at least as far as the energies are concerned, see
figure~\ref{fig:allp03}.%
\begin{figure}
\begin{center}\includegraphics[width=10cm,keepaspectratio,clip]{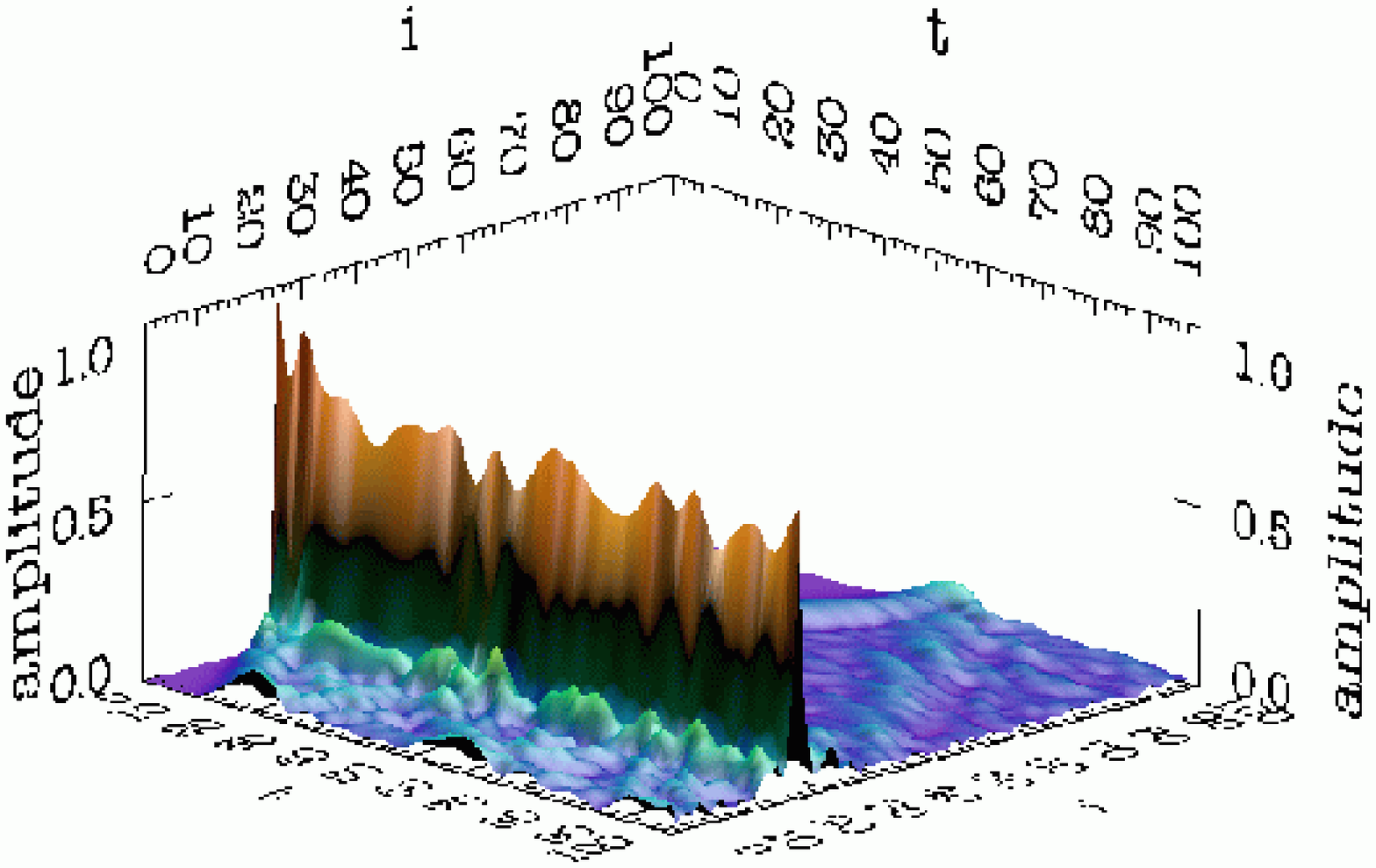} \includegraphics[width=10cm,keepaspectratio,clip]{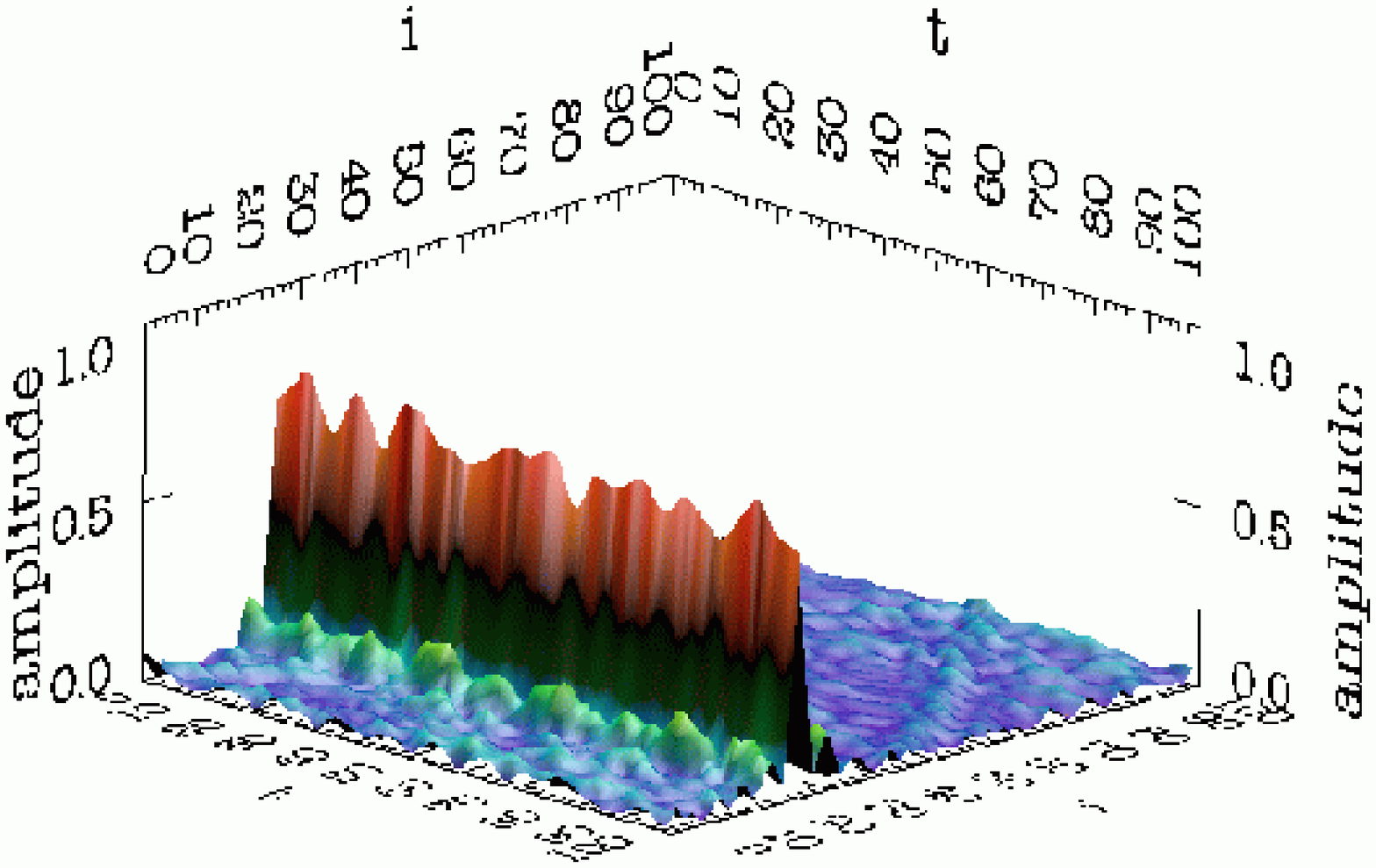}\end{center}

\caption{\label{fig:normp03} Time evolution of the occupation of the site as a function
of the position along the backbone (sum over both) for $J=0.03$.  $t$
gives a time slice number (100 slices used for each figure).  The top
figure shows the early time evolution, the lower one the late time
steady state breather solution. A dynamical representation of the
evolution is available at http://walet.phy.umist.ac.uk/DNA/Jp03.php.}
\end{figure}

The observed time evolution of the amplitude shows that we do have a
stationary solution, but the effect of the small part of the
probability density that propagates is much more pronounced, and makes
things look much more chaotic (figure~\ref{fig:normp03}). In the early
time picture we can also see that these probability waves get
reflected back towards the soliton (and also get reflected by the
soliton).  Clearly, the interesting dynamics observed here deserves
further study.

\begin{figure}
\begin{center}\includegraphics[width=10cm,keepaspectratio,clip]{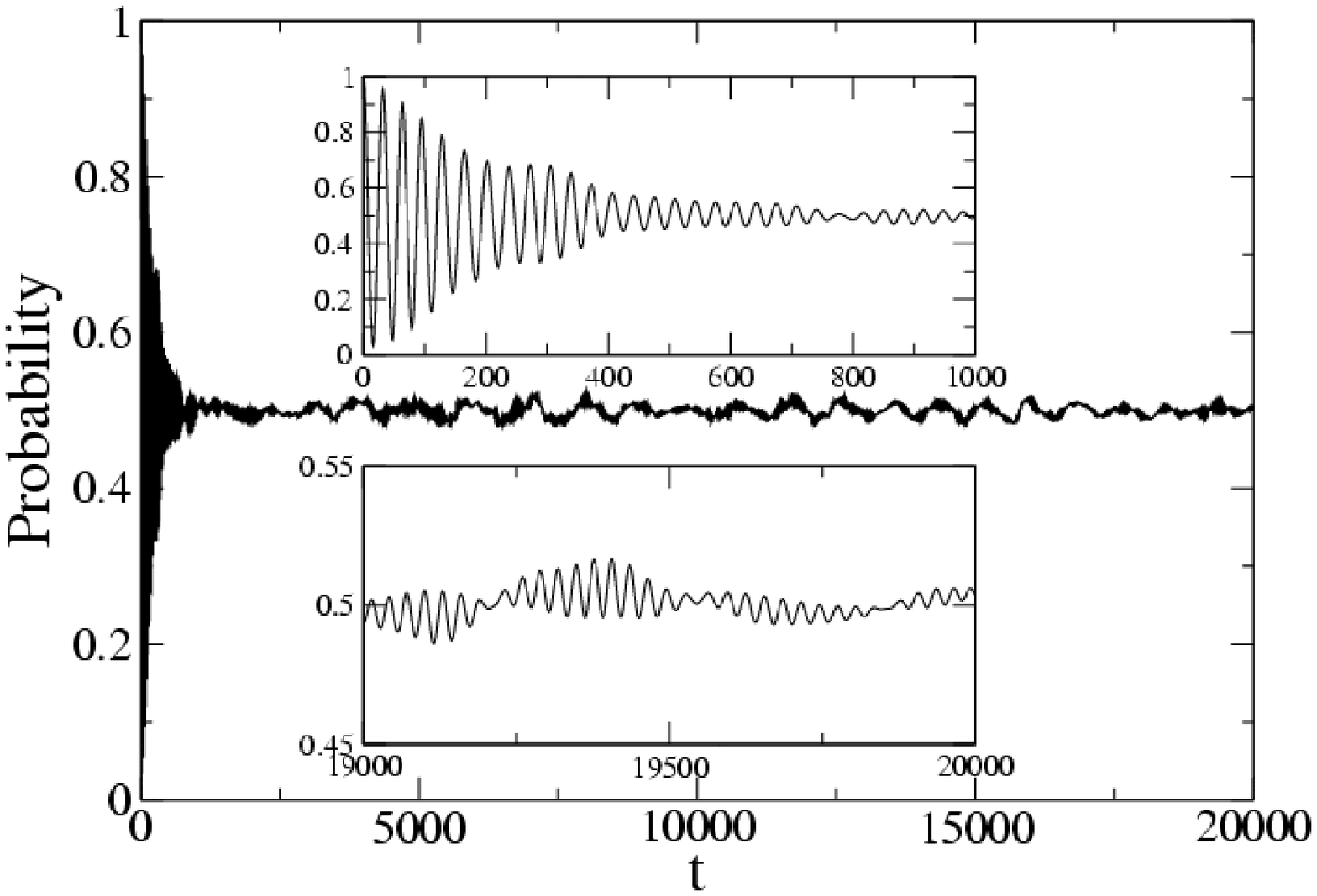}\end{center}

\caption{\label{fig:normbackbonep03} Time evolution of the total occupation
of one of the backbones as a function of time for $J=0.03$. }
\end{figure}

We also note similar structure in the oscillations from one backbone
to the next, see figure~\ref{fig:normbackbonep03}. The oscillations are
more pronounced, and their envelope is probably due to the probability
bouncing from the boundaries. The fastest oscillation is then, almost
certainly, the frequency of the breather.%
\begin{figure}
\begin{center}\includegraphics[width=10cm,keepaspectratio,clip]{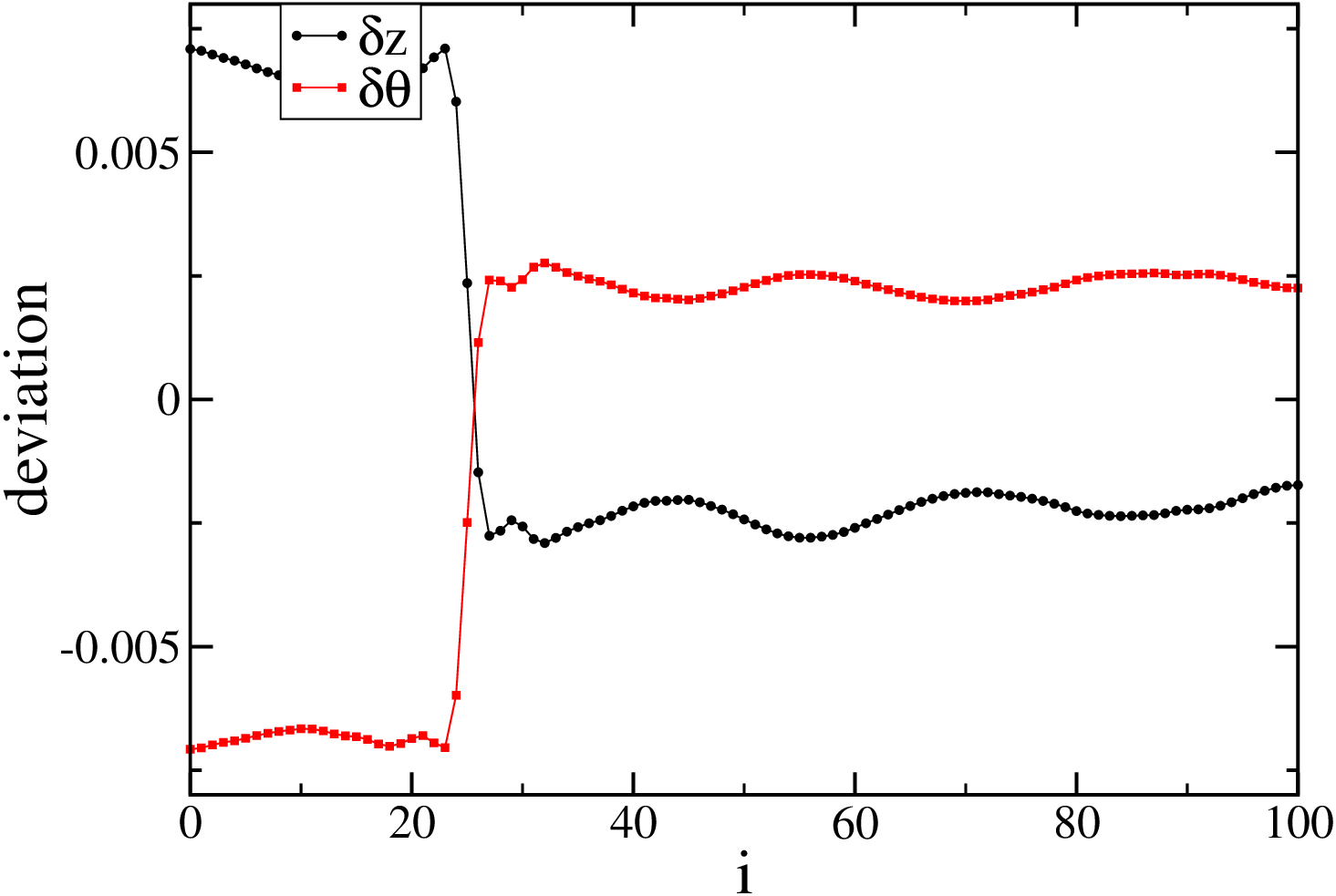}\end{center}

\caption{\label{fig:xzp03} Deviation of $\theta$ and $z$ from their equilibrium
values for the final simulation time, $t=19395$, as a function of the
lattice point label $i$ and $J=0.03$. }
\end{figure}

The sharp jump of $z$ and $\theta$ at the soliton position can again
be seen in figure~\ref{fig:xzp03}.

\begin{figure}
\begin{center}\includegraphics[width=10cm,keepaspectratio,clip]{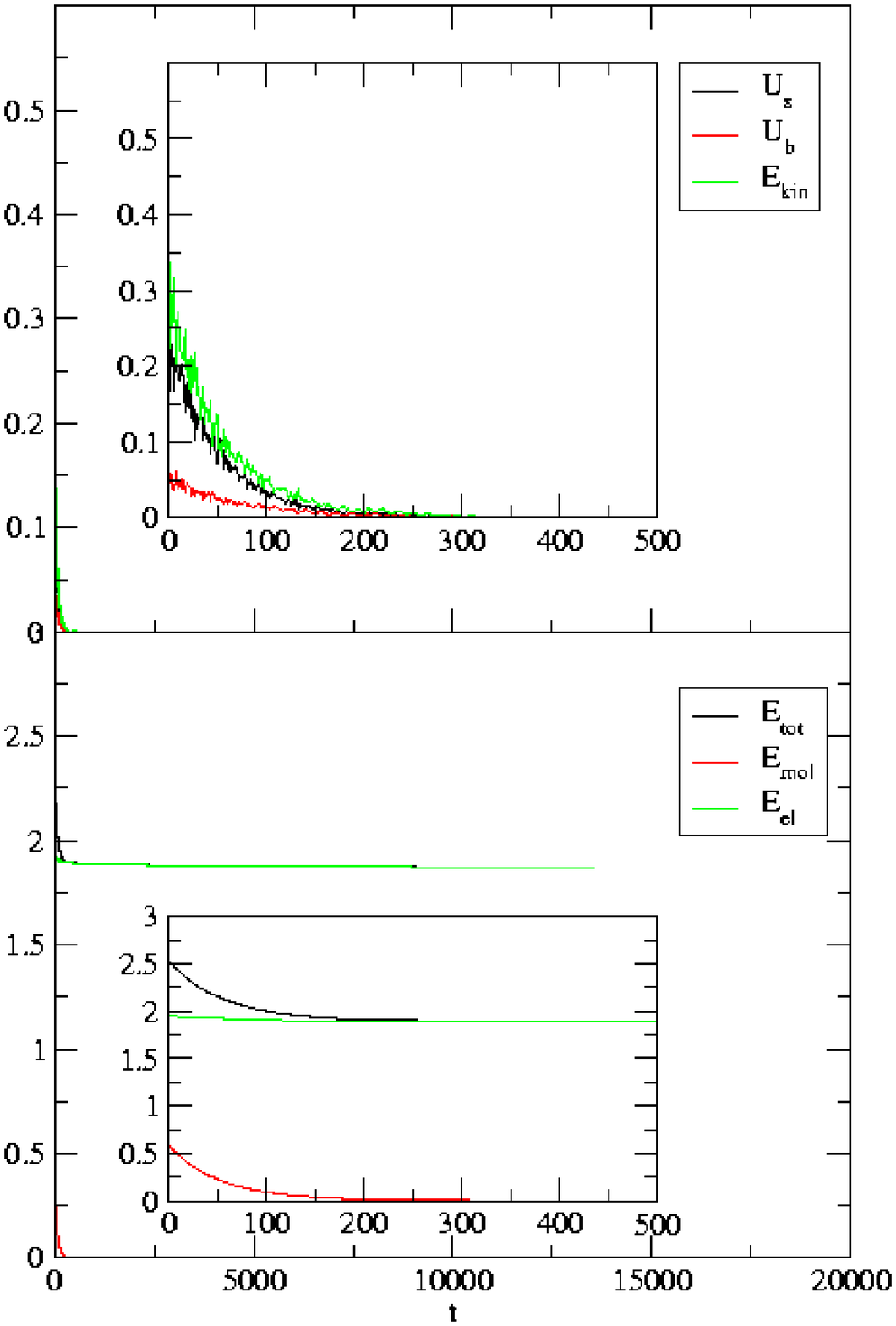}\end{center}

\caption{\label{fig:allp04} Energetics of the relaxation of a 101-rung chain
for $J=0.04$. The upper panel with inset shows the decay of the
mechanical energy, the lower panel the decay of the total and
electronic energies}
\end{figure}

The situation changes drastically if we increase $J$ to $0.04$.  We
still have a well defined relaxation, figure~\ref{fig:allp04}, but there
is no longer a stable soliton.

\begin{figure}
\begin{center}\includegraphics[width=10cm,keepaspectratio,clip]{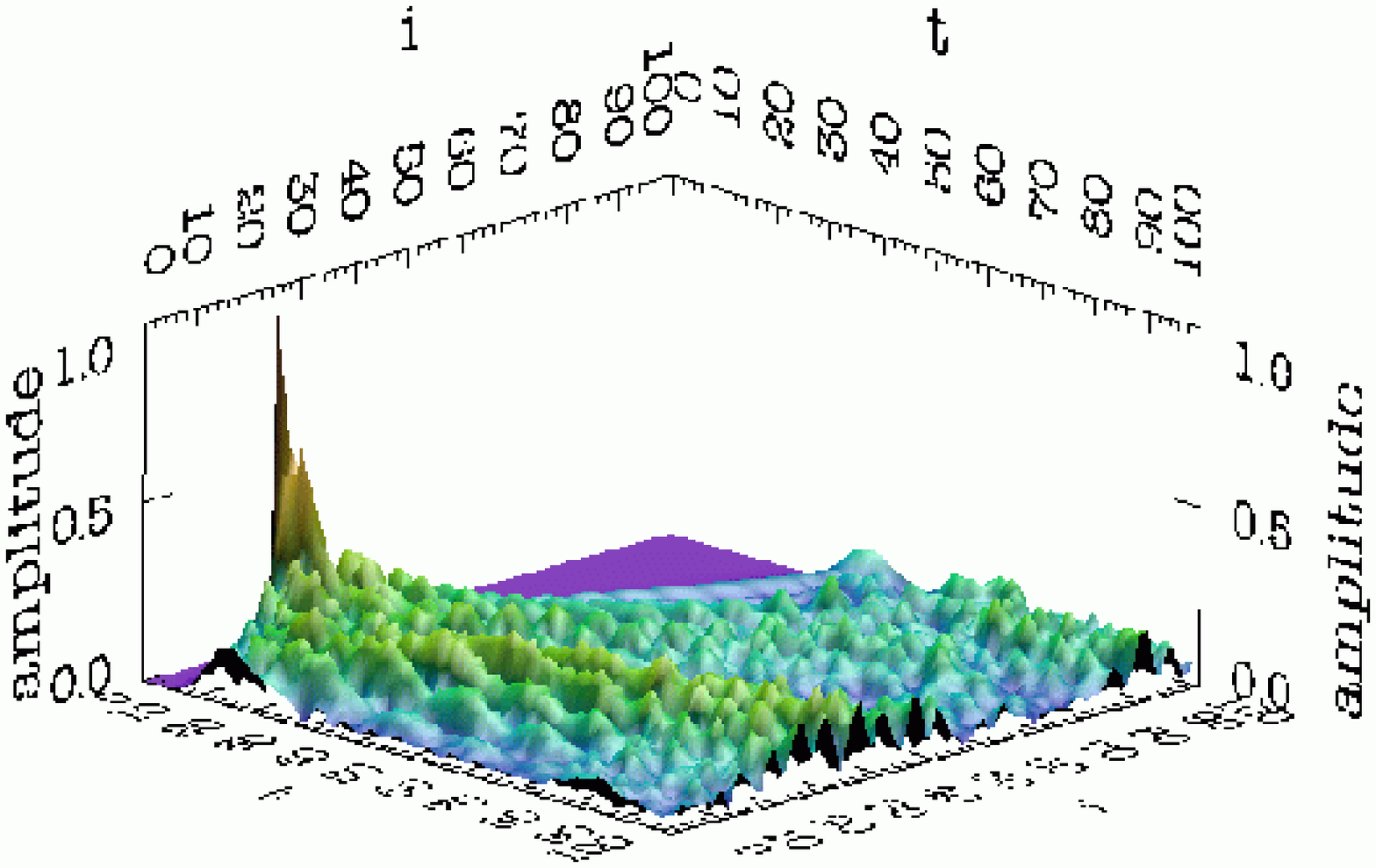} \includegraphics[width=10cm,keepaspectratio,clip]{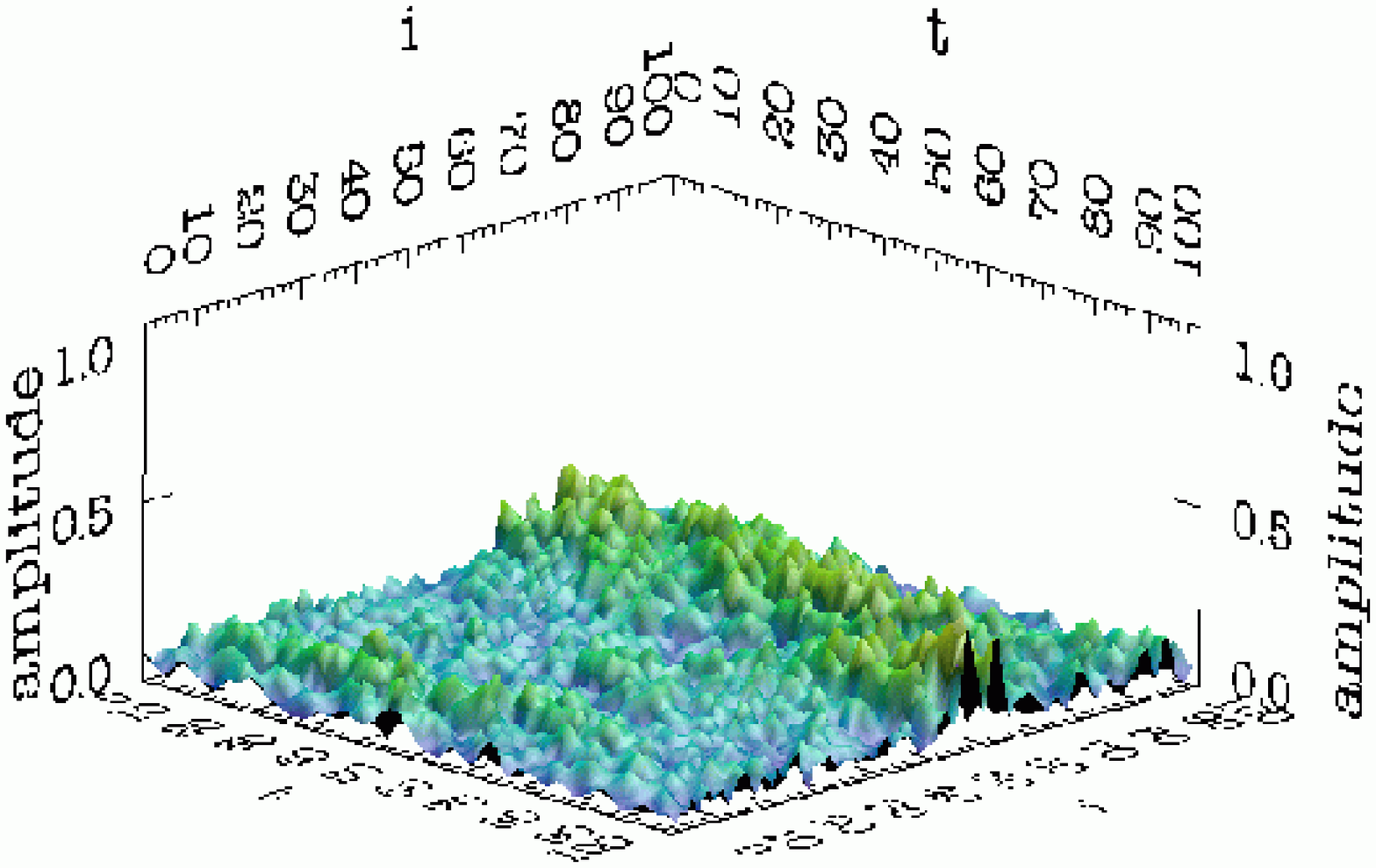}\end{center}

\caption{\label{fig:normp04} Time evolution of the occupation as a
function of the position along the backbone (summed over both) for
$J=0.04$.  $t$ gives a time slice number (100 slices used for each
figure).  The top figure shows the early time evolution, the lower one
the late time {}``chaos''.  A dynamical representation of the
evolution is available at http://walet.phy.umist.ac.uk/DNA/Jp04.php. }
\end{figure}

In figure~\ref{fig:normp04} we demonstrate this. We note a fast decay of
the soliton, even before the reflected waves have a chance to return
to it. This suggest the absence of a stable breather. At later times,
as shown in the lower panel, we see that a soliton-like structure gets
reassembled at a different position.  This may well indicate that at
this point the solitons are only marginally unstable.

\begin{figure}
\begin{center}\includegraphics[width=10cm,keepaspectratio,clip]{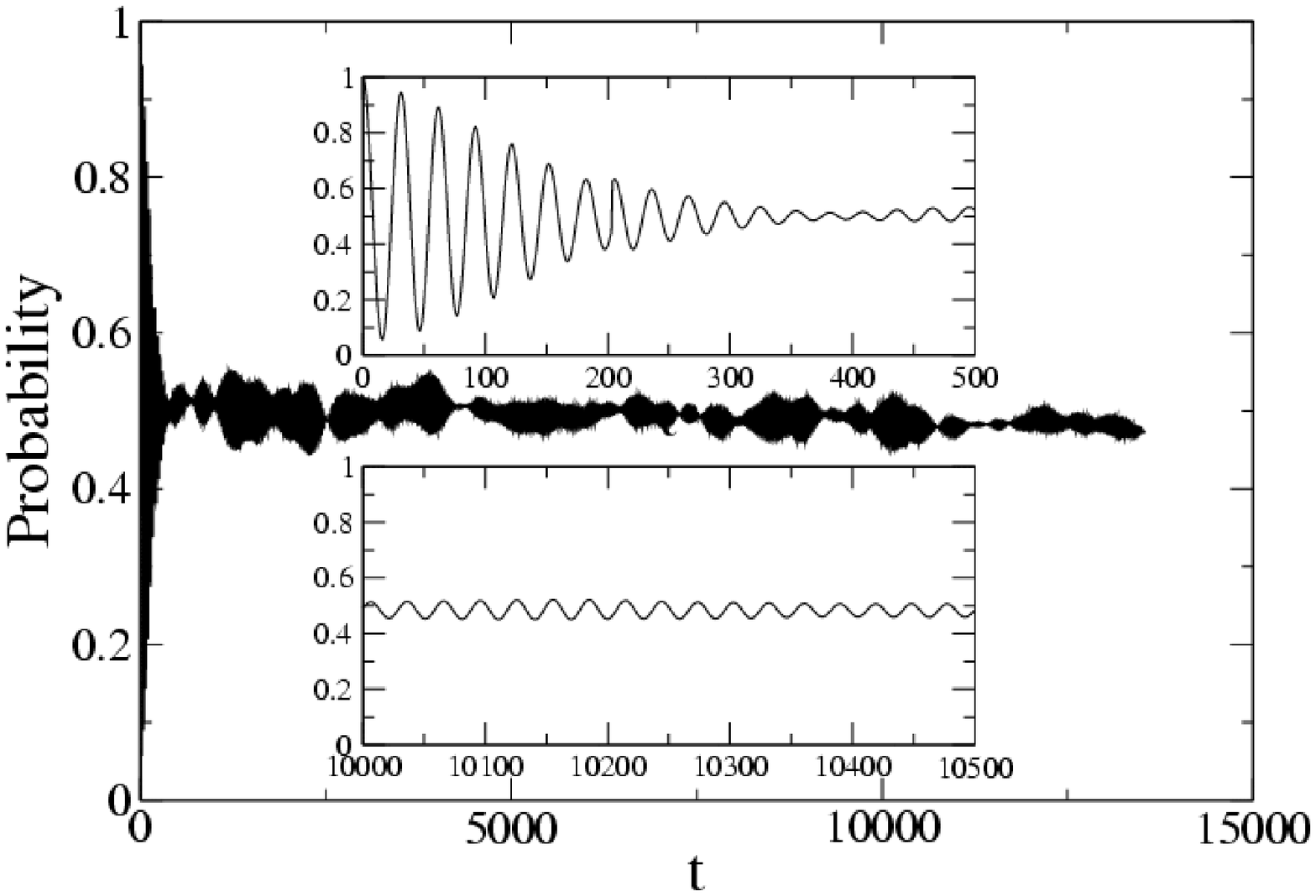}\end{center}

\caption{\label{fig:normbackbonep04} Time evolution of the total occupation
of one of the backbones as a function of time for $J=0.04$. }
\end{figure}

The envelope of the norm also shows more structure, even though this
structure seems to be overlaid on a very regular pattern,
figure~\ref{fig:normbackbonep04}.%

\begin{figure}
\begin{center}\includegraphics[width=10cm,keepaspectratio,clip]{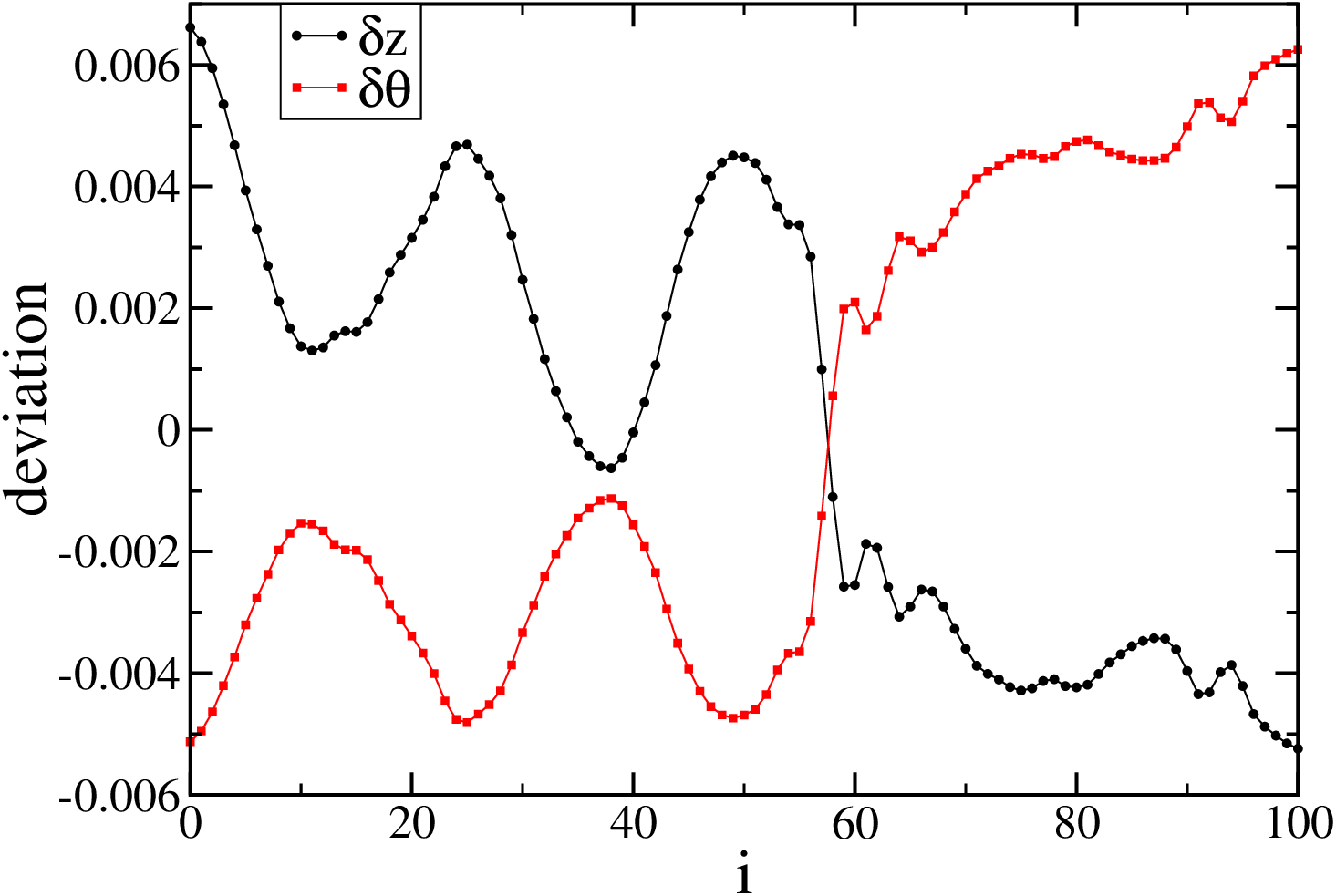}\end{center}

\caption{\label{fig:xzp04} Deviation of $\theta$ and $z$ from their equilibrium
values for the final simulation time, $t=13339.5$, as a function of
the lattice point label $i$ and $J=0.04$. }
\end{figure}

The coupling to the position of the mechanical structure of the double
helix is substantial, as can be seen from the results shown in
figure~\ref{fig:xzp04}.%

\section{Conclusions}

We have constructed a model that describes the chiral movement
of electrons in a double helix. We have shown how it describes
both localised and delocalised charge, depending on parameters.

We have shown that the dynamics of a chiral molecule chain can be
quite involved and interesting, leading to a complicated interplay of
stretching and twisting modes of the underlying double helix generated
by the movement of the electrons. As $J_0$ increases the system
exhibits a sharp transition from localised solitons (charge density)
to a completely delocalised system with a complicated temporal
structure. In each case we see an anticorrelation between $z$ and
$\theta$, with the soliton deforming the lattice by locally reducing
the spacing of the double helix and increasing its twist.

Our  next step will be to try to extend these calculations to charge
transport, where electrons come in on one backbone at one end, and
leave through the same or the other backbone at the other side.

\ack The authors acknowledge support by the EPSRC under grants
GR/N28320/01 (WJZ) and GR/N15672 (NRW).  One of the authors (NRW) is
grateful for hospitality at the Department of Mathematical Sciences,
University of Durham, which contributed greatly to this paper.



\end{document}